\documentclass[journal]{IEEEtran}

\usepackage{cite}
\usepackage{dblfloatfix}
\usepackage{mathtools}
\DeclarePairedDelimiter{\abs}{\lvert}{\rvert} 
\usepackage{float}
\usepackage{amsmath,amssymb,amsfonts}
\usepackage{algorithm}
\usepackage[noend]{algpseudocode}
\usepackage{graphicx}
\graphicspath{{./Figures/}}
\usepackage{textcomp}
\usepackage{comment}
\usepackage{gensymb} 
\usepackage{booktabs}
\usepackage{siunitx}
\usepackage{bm} 
\usepackage{cite} 
\usepackage{cleveref} 
\usepackage{url}
\usepackage{array,multirow}
\usepackage[table]{xcolor}
\usepackage{xcolor,colortbl}
\def\BibTeX{{\rm B\kern-.05em{\sc i\kern-.025em b}\kern-.08em
T\kern-.1667em\lower.7ex\hbox{E}\kern-.125emX}}

\begin{document}

%


\title{A Unified Power-Setpoint Tracking Algorithm for Utility-Scale PV Systems with Power Reserves and Fast Frequency Response Capabilities}

%
%
%

\author{Victor~Paduani,~\IEEEmembership{Student Member,~IEEE,}
        Hui~Yu,~\IEEEmembership{Member,~IEEE,}
        Bei~Xu,~\IEEEmembership{Student Member,~IEEE,}
        and~Ning~Lu,~\IEEEmembership{Fellow,~IEEE}
\thanks{This research is supported by the U.S. Department of Energy's Office of Energy Efficiency and Renewable Energy (EERE) under the Solar Energy Technologies Office Award Number DE-EE0008770. 
Authors names are with the Department
of Electrical and Computer Engineering, North Carolina State University, Raleigh,
NC, 27606 USA e-mail: (vdaldeg@ncsu.edu).}
\thanks{Manuscript received April xx, 2021; revised August xx, 2021.}
}

\maketitle

\begin{abstract}
This paper presents a fast power-setpoint tracking algorithm to enable utility-scale photovoltaic (PV) systems to provide high quality grid services such as power reserves and fast frequency response. The algorithm unites maximum power-point estimation (MPPE) with flexible power-point tracking (FPPT) control to improve the performance of both algorithms, achieving fast and accurate PV power-setpoint tracking even under rapid solar irradiance changes. The MPPE is developed using a real-time, nonlinear curve-fitting approach based on the Levenberg-Marquardt algorithm. A modified adaptive FPPT based on the Perturb and Observe technique is developed for the power-setpoint tracking. By using MPPE to decouple the impact of irradiance changes on the measured PV output power, we develop a fast convergence technique for tracking power-reference changes within three FPPT iterations. Furthermore, to limit the maximum output power ripple, a new design is introduced for the steady-state voltage step size of the adaptive FPPT. The proposed algorithm is implemented on a testbed consisting of a 500 kVA three-phase, single-stage,
utility-scale PV system on the OPAL-RT eMEGASIM platform. Results show that the proposed method outperforms the state-of-the-art.
\end{abstract}


\begin{IEEEkeywords}
\textit{Inverter control, fast frequency response, FPPT, MPPE, power curtailment, power regulation, power reserves, PV system.}
\end{IEEEkeywords}

%
\IEEEpeerreviewmaketitle

\section{Introduction}
%
%
%
%
\IEEEPARstart{B}{ecause} the penetration of photovoltaic (PV) systems is increasing rapidly, grid standards now require advanced grid-support functionalities from utility-scale PV farms. For instance, both the IEEE Standard 1547-2018 \cite{ieee1547ieee} and the California Electric Rule no. 21 \cite{california2016electric} enforce PV inverters to provide active power curtailment, frequency-watt/volt-var droop, and disturbance ride-through capabilities.  

A PV system is able to curtail its output power to follow power setpoints by regulating the voltage applied to its solar panels \cite{ishaque2012improved}, \cite{hoke2013active}. Thus, to track the setpoint accurately while solar irradiance is constantly changing, two major functions are needed: accurately predict the voltage reference corresponding to the power setpoint \cite{hoke2017rapid} and estimate the maximum power point (MPP) available. The later is especially important if the PV system is operated in the power curtailment mode \cite{blanes2012site}. If both functionalities can be achieved, the PV system can be controlled similarly to a battery energy storage system, where the PV can be operated below its MPP in order to maintain a headroom for providing power reserve.

Although many maximum power point tracking (MPPT) algorithms \cite{de2012evaluation,soon2014fast,ahmed2018enhanced} have been proposed in the literature to enable PV systems to track their peak available power, the research on developing power setpoint tracking algorithms for PV curtailment is rather recent. One of the first evaluations of power curtailment methods,  denominated constant power generation or flexible power point tracking (FPPT) in the literature, was presented by Sangwongwanich \textit{et al.} in \cite{sangwongwanich2017benchmarking}. The authors demonstrated that the voltage tracking could be achieved either with PI controllers or based on the Perturb \& Observe (P\&O) technique. An extensive comparison of eleven of the most prominent FPPT methods published in literature was introduced by Tafti \textit{et al.} in \cite{tafti2020extended}, and it concluded that the highest robustness was found in P\&O-based methods. 

The main disadvantage of P\&O-based methods is the trade-off between improving the tracking speed and minimizing power ripples in steady-state operation, both of which are related with the selection of the voltage step size at each iteration. Ideally, a large step should be used during transients, whereas a small step should be used in steady-state. In \cite{tafti2018adaptive}, an adaptive  P\&O-based FPPT method was presented with two major contributions: introduced a method to classify the operation between steady-state and transient modes, and proposed the use of a variable step for improved tracking performance. The study in \cite{tafti2020extended} shows that the adaptive FPPT is preferred for power curtailment applications due to its robustness and superior transient performance. However, the algorithm does not provide maximum power point estimation (MPPE), which is required when providing power reserves. Moreover, the algorithm is susceptible to overshoots as demonstrated in \cite{paduani2020maximum}. 

The MPPE for curtailed PV systems has drawn an increasing attention in recent years. Gevorgian \textit{et al.} pointed out in \cite{gevorgian2016advanced} that using irradiance sensors and calibration factors for MPPE is susceptible to significant errors. In \cite{sangwongwanich2017delta} and \cite{gevorgian2019highly}, the authors suggest that at least one PV inverter should be operated at the MPP so its output can be used as a reference for the other inverters. The method in \cite{gevorgian2019highly} further developed a rotation system to decide which inverter should be selected next as the reference. Yet, the idea is only applicable to large PV farms with identical subarrays. Real-time curve-fitting is proposed by Batzelis \textit{et al.} in \cite{batzelis2017power} for providing very accurate MPPE at the expense of high computational efforts. The method was recently expanded in \cite{batzelis2020mpp} to cope with the fast irradiance intermittency. However, the method requires the introduction of an external ripple for proper convergence, and its setpoint tracking method is based on the PI power controller, which cannot provide fast and robust power-setpoint tracking \cite{tafti2020extended}. 


So far, not much work has been done to integrate the MPPE into PV power curtailment to achieve better performance for both algorithms. Zhu \textit{et al.} in \cite{zhu2020high} and Li \textit{et al.} in \cite{li2018novel} developed power-setpoint tracking algorithms including MPPE, but the control scheme only functions in a two-stage PV system that operates on the left side of the MPP. This greatly limits the adoption of the algorithm in practice.

Therefore, this paper presents an algorithm that combines a modified robust P\&O-based FPPT technique with a real-time curve-fitting-based MPPE that can be applied to both single-stage and two-stage topologies. 
In this unified method, the oscillatory nature of the P\&O is leveraged to provide a good measurement window for the curve-fitting, hence discarding the need for an external ripple. Furthermore, in this model the estimation algorithm's results are used not only to provide power reserves but also to improve the dynamic performance of the power setpoint tracking, enabling the PV system to provide fast frequency response for enhanced grid-support. One major advantage of this work compared to another well-known fast frequency response algorithm in the literature \cite{hoke2017rapid} is that it does not require irradiance or PV cell temperature sensors as inputs. The main contributions of the work to the literature are summarized as: 
\begin{itemize}
     \item A unified power curtailment algorithm combining the robust setpoint-tracking performance from P\&O-based FPPT methods with an accurate power reserves estimation via real-time curve-fitting that can be used on both single and two-stage systems.
    \item A fast-convergence technique to adjust the PV power reserves within three iterations of the setpoint tracking algorithm.
    \item A mechanism to decouple the impact of irradiance changes from the iterations of the P\&O technique, so that the relation between PV voltage and power is properly measured when under irradiance intermittency.
    \item A straightforward new design for the steady-state step of the adaptive FPPT to limit the output power ripple.
\end{itemize}


The work is divided as follows. In Section II, the modeling and the major functionalities of the proposed algorithm are explained. Section III presents the simulation results, and Section IV concludes the work with final remarks. 

\section{Modeling and Control Methodologies}

The circuit and control system of a single-stage, three-phase PV system is displayed in Fig. 1. A hierarchical control structure composed of a dc-link voltage controller cascaded with a current controller is used to generate the inverter modulation signal, $\vec{m}$. More details regarding the generation of the modulation signal and the PV array model utilized can be found in \cite{paduani2020maximum}. When the PV system is required to follow a power setpoint to provide grid services (e.g., provide spinning reserves or regulation up or down services), the FPPT control block will require inputs from the MPPE block to calculate the voltage reference, $V_{\mathrm{dc}}^{*}$.

In the next subsections, we will first present the PV model and the implemented MPPE, next introduce the new design for the voltage step size of the power reference tracking algorithm at steady-state, and then present two new control techniques: irradiance decoupling and rapid setpoint tracking, the first for achieving better P\&O performance, and the second for a fast convergence to the targeted power reference to enable fast frequency response.


\subsection{Modeling of the PV Array}

A PV array can be represented by the single-diode model introduced in \cite{villalva2009comprehensive} as shown in Fig. \ref{Fig:pvcircuit}. The model can be described by the following equations as presented in \cite{batzelis2017simple}.

\begin{figure*}[!t]
	\centering
	\includegraphics[width=1\textwidth]{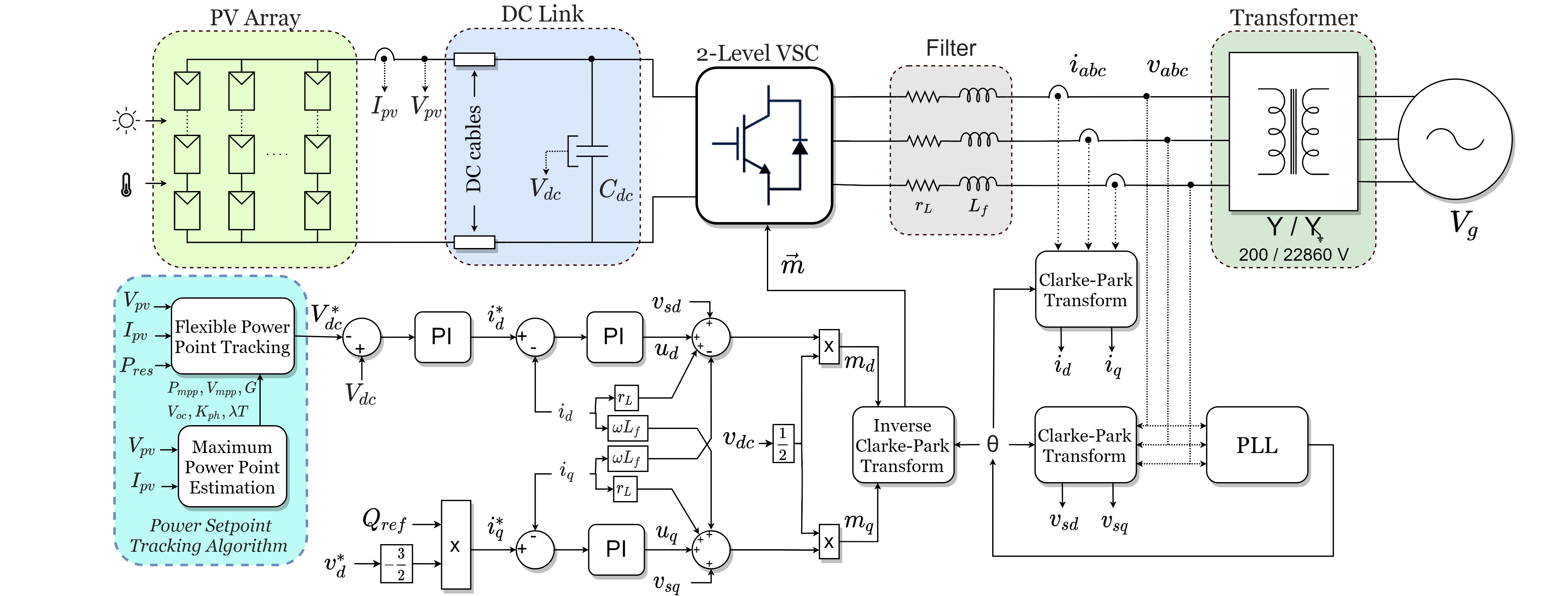}
	\caption{Circuit and control system block diagrams of a utility-scale PV system.}
	\label{maindiagram}
\end{figure*}

\begin{figure}[htb]
	\centerline{\includegraphics[width=0.3\textwidth]{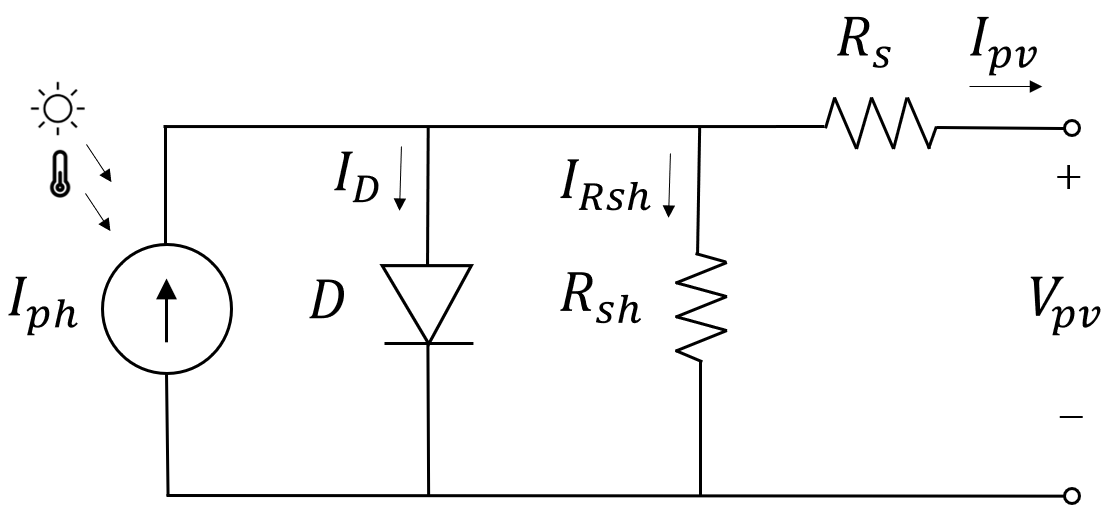}}
	\caption{Single-diode PV model circuit.}
	\label{Fig:pvcircuit}
\end{figure} 

\begin{equation}
	I_{\mathrm{pv}}= I_{\mathrm{ph}} - I_{\mathrm{D}} - I_{\mathrm{Rsh}}
	\label{PVeq2}
\end{equation}

\begin{equation}
	I_{\mathrm{pv}}= I_{\mathrm{ph}} - I_{\mathrm{s}}\Big( e^{\frac{V_{\mathrm{pv}}+I_{\mathrm{pv}}R_{\mathrm{s}}}{a}}-1\Big) - \Big(\frac{V_{\mathrm{pv}}+I_{\mathrm{pv}}R_{\mathrm{s}}}{R_{\mathrm{sh}}}\Big)
	\label{PVeq}
\end{equation}

\begin{equation}
	a = \delta_{0} V_{\mathrm{oc0}}\lambda T
	\label{eq:a}
\end{equation}

\begin{equation}
	R_{\mathrm{s}}= [a_{0}(w_{0}-1)-V_{\mathrm{mp0}}]/I_{\mathrm{mp0}}
	\label{Rs}
\end{equation}

\begin{equation}
	R_{\mathrm{sh}}= \delta_{0} V_{\mathrm{oc0}}(w_{0}-1)/[I_{\mathrm{sc0}}(1-1/w_{0})-I_{\mathrm{mp0}}]/G
	\label{Rsh}
\end{equation}

\begin{equation}
	I_{\mathrm{ph}}= (1+R_{\mathrm{s0}}/R_{\mathrm{sh0}})I_{\mathrm{sc0}}G[1+\alpha_{\mathrm{Isc}}T_{0}(\lambda T-1)]
	\label{eq:Iph}
\end{equation}

\begin{equation}
	I_{\mathrm{s}}= I_{\mathrm{ph0}}e^{-1/\delta_{0}}\lambda T^{3}e^{47.1(1-1/\lambda T)}
	\label{eq:Is}
\end{equation}

\begin{equation}
	\delta_{0}= (1 - \beta_{\mathrm{Voc}}T_{0})/(50.1 - \alpha_{\mathrm{Isc}}T_{0})
	\label{eq:d0}
\end{equation}

\begin{equation}
	 w_{0} = W\big\{e^{1/\delta_{0} +1}\big\}
	 \label{eq:d0-w0}
\end{equation}

\begin{equation}
    a_{0} = \delta_{0}V_{\mathrm{oc0}}
    \label{eq:a0}
\end{equation}


\begin{equation}
    R_{\mathrm{sh0}} = a_{0}(w_{0}-1)/[I_{\mathrm{sc0}}(1-1/w_{0})-I_{\mathrm{mp0}}]
    \label{eq:Rsh0}
\end{equation}

\begin{equation}
    I_{\mathrm{ph0}} = (1+R_{\mathrm{s0}}/R_{\mathrm{sh0}})I_{\mathrm{sc0}}
    \label{I_{ph0}}
\end{equation}

\begin{equation}
    I_{s0} = I_{\mathrm{ph0}}e^{-1/\delta_{0}}
    \label{eq:Is0}
\end{equation}

By applying KCL to the circuit, the PV output current is defined as (\ref{PVeq}). The five parameters of the model are the ideality factor ($a$), the series resistance ($R_{\mathrm{s}}$), the shunt resistance ($R_{\mathrm{sh}}$), the photocurrent ($I_{\mathrm{ph}}$), and the diode saturation current ($I_{\mathrm{s}}$). The five PV model parameters, $a_{0}$, $R_{\mathrm{s0}}$, $R_{\mathrm{sh0}}$, $I_{\mathrm{ph0}}$, and $I_{\mathrm{s0}}$, are the baseline values calculated under the standard test conditions (STC). The equation for the series resistance at STC ($R_{s0}$) is omitted since $R_{\mathrm{s}} = R_{\mathrm{s0}}$. From the manufacturer datasheet or measured in field at STC, the short-circuit current ($I_{\mathrm{sc0}}$), the open-circuit voltage ($V_{oc0}$), and the voltage and current at MPP ($V_{\mathrm{mp0}}$, $I_{\mathrm{mp0}}$) can be obtained to calculate the five baseline parameters at STC and used in (10)-(13). After that, the five parameters at any irradiance and temperature can be calculated using (3)-(7). Note that $G$ is the normalized irradiance, and $\lambda T$ is the ratio between the cell temperature and its temperature at STC ($T_{0}$), corresponding to 298.15 K. Moreover, $\delta_{0}$, given by (\ref{eq:d0}), is a ratio between the ideality factor and the open circuit voltage (V$_{\mathrm{oc}}$), whereas $w_{0}$ is an auxiliary parameter related to $\delta$ via the Lambert $W$ function (\ref{eq:d0-w0}). In this work, the Lambert $W$ function is solved with the method from \cite{moler2021lambert}. In addition, $\beta_{\mathrm{Voc}}$ and $\alpha_{\mathrm{Isc}}$, in (\ref{eq:d0}), correspond to the temperature coefficient of the open-circuit voltage and short-circuit current, respectively.

\subsection{Maximum Power Point Estimation}
In this section we present a modified MPPE algorithm based on the methods introduced in \cite{batzelis2017power} and \cite{batzelis2020mpp}. By inserting (\ref{eq:a})-(\ref{eq:Is}) in (\ref{PVeq}), we obtain (\ref{combinedeq}), which correlates the output voltage and current of a PV system with irradiance and temperature values. Then, by applying a non-linear least squares method to (\ref{combinedeq}), it is possible to estimate the incident irradiance and temperature on the PV array from voltage and current measurements. In \cite{batzelis2017power}, good convergence was obtained via Levenberg-Marquardt (LM). Thus, the same algorithm is utilized in this work.

\small
\begin{multline}
\hspace*{-1em}	a_{0} \lambda T \mathrm{ln} \Bigg[\frac{G I_{\mathrm{ph0}}\big[1 + \alpha_{\mathrm{Isc}} T_{0} (\lambda T-1)\big] - I_{\mathrm{pv}} - G\big(V_{\mathrm{pv}}+\tfrac{I_{\mathrm{pv}} R_{\mathrm{s0}}}{R_{\mathrm{sh0}}}\big)}{I_{\mathrm{s0}}\lambda T^{3}e^{47.1(1-1/\lambda T)}}\Bigg]\\
	-V_{\mathrm{pv}} -I_{\mathrm{pv}}R_{\mathrm{s0}} = 0
	\label{combinedeq}
\end{multline}
\normalsize

The LM algorithm utilizes the damping parameter ($\eta$) to adaptively vary the parameter updates between the gradient descent and the Gauss-Newton methods \cite{gavin2019levenberg}. In (\ref{eq:std_LM}), $\bm{W}$ is the weighting matrix, ($\bm{y} - \hat{\bm{y}}$) is the error between the measured data and the curve-fit function, defined by (\ref{combinedeq}), $\bm{J} = [\tfrac{\partial y}{\partial G} \hspace{0.5em} \tfrac{\partial y}{\partial \lambda T}]^T$ is the Jacobian, and $\bm{\delta}$ is the vector of parameters updates in each LM iteration. For each ($V_{\mathrm{pv}}$, $I_{\mathrm{pv}}$) measurements added to the measurement window, the partial derivatives from the Jacobian are calculated. Once the measurement window is full, one iteration of the LM is performed with (\ref{LM2}) and (17).

\begin{equation}
	{\Big[\bm{J}^{T} \bm{W J} + \eta \,\mathrm{diag} \big(\bm{J}^T \bm{W J}\big) \Big]}\bm{\delta} = \bm{J^{T} W} (\bm{y} - \hat{\bm{y}})
	\label{eq:std_LM}
 \end{equation}
 
 \begingroup
\renewcommand*{\arraystretch}{1.5}

\begin{equation}
\bm{\Omega_{i}} = \begin{bmatrix} (1+\eta)\sum\limits_{n=1}^{N}\tfrac{\partial^2 y_{\mathrm{n}}}{\partial \lambda T^2} & \sum\limits_{n=1}^{N}\tfrac{\partial y_{\mathrm{n}}}{\partial G}\tfrac{\partial y_{\mathrm{n}}}{\partial \lambda T}\\ \sum\limits_{n=1}^{N}\tfrac{\partial y_{\mathrm{n}}}{\partial \lambda T}\tfrac{\partial y_{\mathrm{n}}}{\partial G} & \sum\limits_{n=1}^{N}(1+\eta)\tfrac{\partial^2 y_{\mathrm{n}}}{\partial G^2}   \end{bmatrix} 
\label{LM2}
\end{equation}
\endgroup

\begin{align}
    \bm{\delta_{i}} = \begin{bmatrix} \Delta G_{i} \\ \Delta \lambda T_{i} \end{bmatrix} = \bm{\Omega_{i}}^{-1}\begin{bmatrix}
\sum\limits_{n=1}^{N}\tfrac{\partial y_{\mathrm{n}}}{\partial G}(y_{\mathrm{n}}- \hat{y_{\mathrm{n}}}) \\ \sum\limits_{n=1}^{N}\tfrac{\partial y_{\mathrm{n}}}{\partial \lambda T}(y_{\mathrm{n}}- \hat{y_{\mathrm{n}}})
\end{bmatrix}  \\  \nonumber|\Delta G_{i}| \leq \Delta G_{\mathrm{max}}, \nonumber|\Delta\lambda T_{i}| \leq \Delta T_{\mathrm{max}}
\end{align}

Because this is a real-time non-linear curve-fitting, irradiance intermittency and measurement noise can impact the algorithm's convergence. Therefore, this work proposes the addition of saturation factors ($\Delta G_{\mathrm{max}}$, $\Delta T_{\mathrm{max}}$) and a maximum irradiance threshold to the parameter updates from each LM iteration. The saturation factors and the irradiance threshold help to maintain the curve-fitting convergence in case of a poor measurement window. The saturation factors are defined based on the maximum expected changes in irradiance and temperature that may be experienced by the solar array, whereas the irradiance threshold is based on the maximum expected irradiance. References \cite{marion2014new}, \cite{EPRIdata}, and \cite{marcos2011irradiance} can be used for finding appropriate values. In this work, the estimated irradiance threshold is set to the nominal value of 1000 W/m$^{2}$. However, this value can be boosted to higher limits such as 1100 or 1200 W/m$^{2}$ to account for edge-of-cloud lens effects.

Here, the following strategy is proposed for the update of the damping parameter ($\eta$). First, the squared residual $(\bm{y_{i}}- \hat{\bm{y_{i}}})^2$ is compared for three different sets of damping values: $a\eta$, $\eta/a$, and $\eta$. If the smallest residual is obtained by updating the parameters with a higher damping value, then the new damping value is set as $a\eta$. If the smallest is obtained by decreasing the damping, then the new damping is set as $\eta/a$. Otherwise, the damping value is maintained. Finding the best strategy for updating the LM damping parameter is a well-researched topic in the literature. The approach is normally heuristic-based. The strategy presented in this paper is similar to approaches found in \cite{gavin2019levenberg}, \cite{nielsen1999damping}, and \cite{long2020diesel}. Besides, note that in this work the MPPE does not need an external ripple for achieving satisfactory curve-fitting performance. This is because the model is combined with a P\&O-based technique, which naturally provides the necessary oscillation around the operating point for a measurement window with enough curve-length coverage.


Due to the high computational effort demanded by the curve-fitting, its period ($T_{\mathrm{LM}}$) must be large (e.g., 2-10 s). Consequently, the irradiance estimation may not provide adequate convergence when irradiance varies rapidly. To address this problem, we adopt an improved real-time curve-fitting technique introduced by Batzelis \textit{et al.} in \cite{batzelis2020mpp}. Assuming that changes in temperature are much slower than changes in irradiance, $\lambda T$ can be considered as a constant between LM iterations. Thus, the irradiance can be directly calculated from voltage and current measurements using (\ref{eq:G}). With the estimated $G$ and $\lambda T$, the PV voltage and current at MPP can be calculated by (\ref{eq:vmpp}) and (\ref{eq:impp}) \cite{batzelis2015method}, in which $w$, given by (\ref{eq:w-Lambert}), is calculated with the Lambert $W$ function. 
\vspace{-0.075cm}
\begin{equation}
	G = \frac{I_{\mathrm{pv}} + I_{\mathrm{s0}}\lambda T^3 e^{47.1\big(1-\tfrac{1}{\lambda T}\big)}e^{\tfrac{V_{\mathrm{pv}}+I_{\mathrm{pv}}R_{\mathrm{s0}}}{a_{0}\lambda T}}}{I_{\mathrm{ph0}}\big[1+\alpha_{\mathrm{Isc}}T_{0}(\lambda T -1)]-\tfrac{V_{\mathrm{pv}}+I_{\mathrm{pv}}R_{\mathrm{s0}}}{R_{\mathrm{sh0}}}}
	\label{eq:G}
\end{equation}

\begin{equation}
	V_{mp} = \bigg(1+\frac{R_{\mathrm{s}}}{R_{\mathrm{sh}}}\bigg)a(w-1) - R_{\mathrm{s}}I_{\mathrm{ph}} \bigg(1-\frac{1}{w}\bigg)
	\label{eq:vmpp}
\end{equation}
\begin{equation}
	I_{\mathrm{mp}} = I_{\mathrm{ph}}\bigg(1-\frac{1}{w}\bigg) - \frac{a(w-1)}{R_{\mathrm{sh}}}
	\label{eq:impp}
\end{equation}

\begin{equation}
	w = W\bigg\{ I_{\mathrm{ph}} \frac{e}{I_{\mathrm{s}}}\bigg\}
	\label{eq:w-Lambert}
\end{equation}

\begin{figure}[htb]
	\centerline{\includegraphics[width=0.5\textwidth]{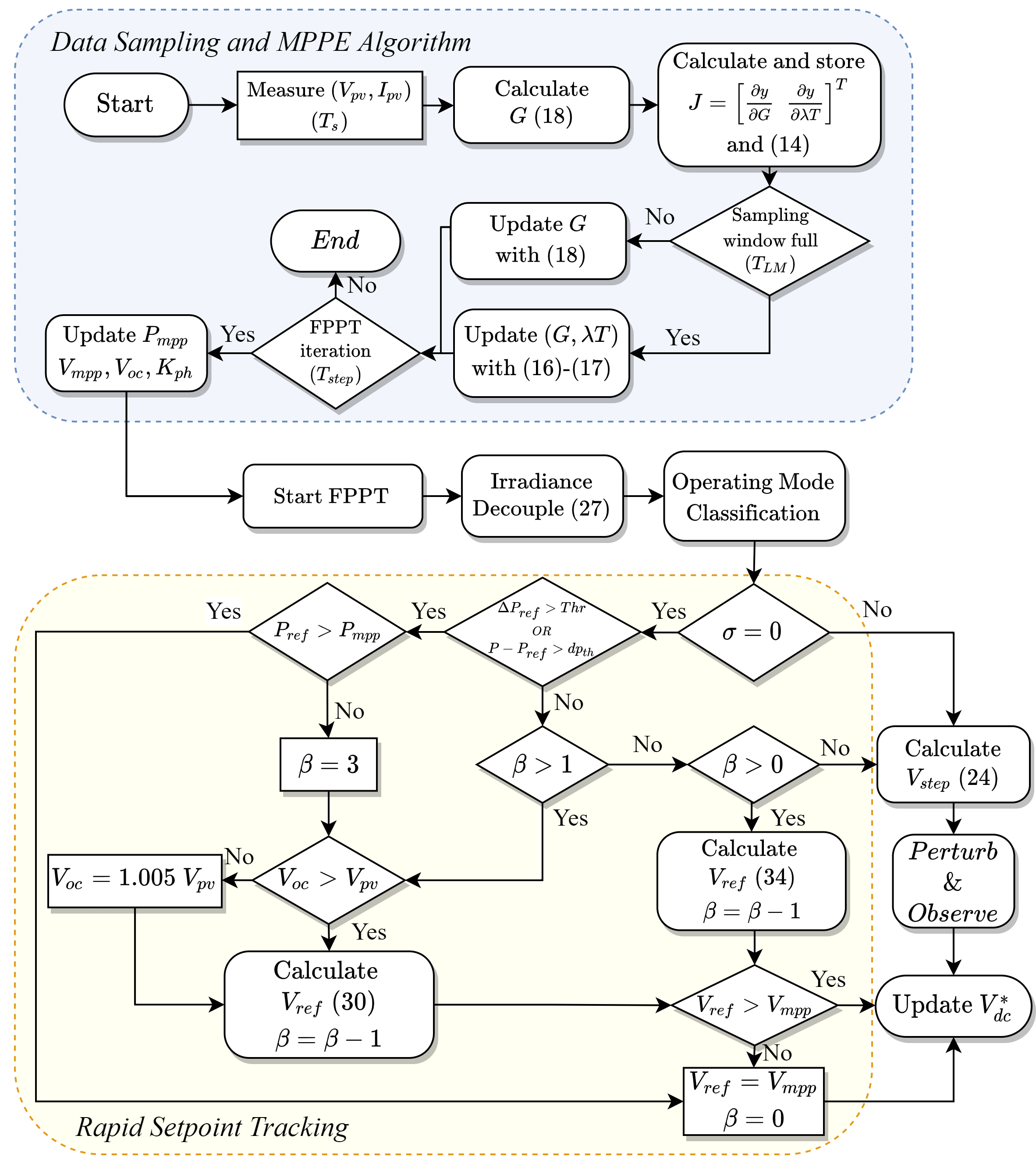}}
	\caption{Flowchart of the unified power curtailment algorithm proposed.}
	\label{flowchart}
\end{figure}

Figure \ref{flowchart} displays a flowchart of the proposed algorithm with the MPPE functionality highlighted in blue. Note that the proposed algorithm has three distinct iterative loops. The fast sampling loop runs every $T_{s}$, in which ($V_{\mathrm{pv}},I_{\mathrm{pv}}$) are sampled and added to a measurement window for calculating $G$, updating the sums from (\ref{LM2}), and computing the residual (\ref{combinedeq}). The medium-speed updating loop runs every $T_{\mathrm{step}}$, in which the voltage reference for power setpoint tracking is updated. The slow updating loop runs every $T_{\mathrm{LM}}$, in which the LM parameters in (17) are updated. In the first LM iteration, the temperature is initialized using its STC value obtained at 25\degree C ($\lambda T = 1$). Note that the slow updating loop is triggered once the window storing ($V_{\mathrm{pv}},I_{\mathrm{pv}}$) measurements is full. It is worth mentioning that the matrix inversion from (17) can be algebraically derived beforehand to alleviate the computational burden. 

Moreover, in each medium-speed updating loop, the irradiance decoupling method proposed in this work and the operating mode classification method introduced in \cite{tafti2018adaptive} are executed. Then, the voltage reference is updated either by the rapid setpoint tracking (RST) technique, or by the P\&O method (i.e., using the $V_{\mathrm{step}}$ introduced in \cite{paduani2020maximum}). The functionalities developed for the medium speed iteration will be explained in detail next.

\subsection{Power Setpoint Tracking}



\begin{figure}[htb]
	\centerline{\includegraphics[width=0.5\textwidth]{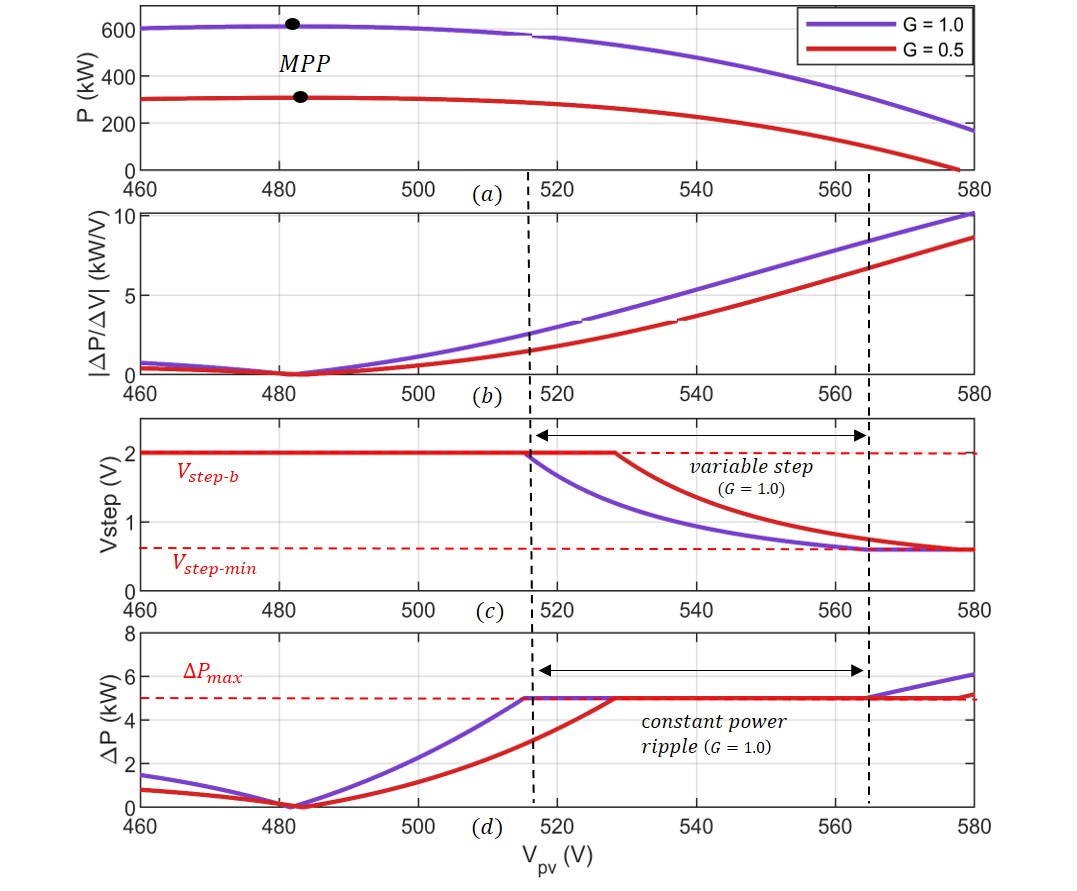}}
	\caption{Illustration of the steady-state  voltage step calculation strategy for a 612 kW PV system: (a) Power-voltage curve, (b) PV slope curve, (c) Steady-state voltage steps, (d) Expected output power ripples when using $V_{\mathrm{step}\text{-}\mathrm{ss}}$ with $\Delta P_{\mathrm{max}}=5 $ kW.}
	\label{Fig:tracking}
\end{figure}

In steady-state operation, the power-reference tracking algorithm normally adopts a small voltage step to reduce the steady-state output power ripple ($\Delta P_{\mathrm{ss}}$). However, the slope of the P-V curve (i.e., $\Delta P/\Delta V$) increases rapidly when the PV operation point passes the MPP and moves towards the right side of the P-V curve, as shown in Figs. \ref{Fig:tracking}(a) and \ref{Fig:tracking}(b). Thus, even if a small steady-state voltage step ($V_{\mathrm{step}\text{-}\mathrm{ss}}$) is selected, the system may still experience a large $\Delta P_{\mathrm{ss}}$ due to the increased $\Delta P/\Delta V$, especially when a large power curtailment is required. To address this issue, in \cite{tafti2018adaptive}, Tafti \textit{et al.} propose to use a variable voltage step, $V_{\mathrm{step}\text{-}\mathrm{ss}}$, which is calculated via a gain applied to the PV slope. This allows $V_{\mathrm{step}\text{-}\mathrm{ss}}$ to be reduced continuously when the operating point moves further along the right side of the PV curve. However, no clear procedure for the design of the gain was proposed in the paper. Moreover, there is no direct control to reduce $\Delta P_{\mathrm{ss}}$, leaving it varying with respect to solar irradiance. Therefore, we propose a new method for calculating $V_{\mathrm{step}\text{-}\mathrm{ss}}$ as 
\begin{align}
	V_{\mathrm{step}\text{-}\mathrm{ss}} = \text{max}\bigg\{\text{min}\Big[\abs*{\tfrac{\Delta V}{\Delta P}}\Delta P_{\mathrm{max}}, V_{\mathrm{step}{\text{-}\mathrm{b}}}\Big],V_{\mathrm{step}{\text{-}\mathrm{min}}}\bigg\}
	\label{eq:vstep_ss}
\end{align}
where $V_{\mathrm{step}\text{-}\mathrm{b}}$ is the base voltage step, $V_{\mathrm{step}\text{-}\mathrm{min}}$ is the minimum voltage step, and $\Delta P_{\mathrm{max}}$ is the maximum acceptable output power ripple.

As illustrated in Figs. 4(c) and 4(d), when $\Delta P_{\mathrm{ss}}$ is below the limit $\Delta P_{\mathrm{max}}$, the step is fixed at its base value $V_{\mathrm{step}\text{-}\mathrm{b}}$. Then, as the voltage increases and the PV slope increases (Fig. 4(b)), $V_{\mathrm{step}\text{-}\mathrm{ss}}$ is reduced as needed to maintain $\Delta P_{\mathrm{ss}}$ at $\Delta P_{\mathrm{max}}$. Note that if the voltage step required to regulate $\Delta P_{\mathrm{ss}}$ is lower than the minimum acceptable value ($V_{\mathrm{step}\text{-}\mathrm{min}}$), $\Delta P_{\mathrm{ss}}$ will no longer be maintained at $\Delta P_{\mathrm{max}}$, causing it to increase as the operation moves further to the right of the PV curve. Therefore, the trade-off between $V_{\mathrm{step}\text{-}\mathrm{min}}$ and $\Delta P_{\mathrm{max}}$ must be accounted for during implementation. 

The main advantage of this method is that it provides a straightforward approach to limit the output power ripple. In contrast, the main disadvantage is that the output power ripple limit may cross $\Delta P_{\mathrm{max}}$ if $V_{\mathrm{step}\text{-}\mathrm{ss}}$ reaches $V_{\mathrm{step}\text{-}\mathrm{min}}$. To maintain control over the output power ripple throughout all ranges of operation,  the desired range of $V_{\mathrm{step}\text{-}\mathrm{ss}}$, the PV array dc-ac power ratio, and the maximum acceptable power ripple at steady-state must be accounted during the design.

During transients, the voltage step is calculated by the method introduced in \cite{paduani2020maximum} as

\begin{align}
	V_{\mathrm{step}\text{-}\mathrm{tr}} = \text{min}\bigg\{K_{\mathrm{tr}}\abs*{P - P_{\mathrm{ref}}}, V_{\mathrm{step}{\text{-}\mathrm{max}}}\bigg\}
	\label{eq:vstep_tr}
\end{align}

\begin{equation}
    V_{\mathrm{step}} = \sigma V_{\mathrm{step}\text{-}\mathrm{ss}} + (1-\sigma)V_{\mathrm{step}\text{-}\mathrm{tr}}
    \label{eq:Vstep}
\end{equation}

\[
    V_{\mathrm{step}} = 
\begin{cases}
    V_{\mathrm{step}\text{-}\mathrm{ss}},& \text{if } \sigma = 1\\
    V_{\mathrm{step}\text{-}\mathrm{tr}},& \text{if } \sigma = 0
    \label{eq:alpha}
\end{cases}
\]

where $V_{\mathrm{step}{\text{-}\mathrm{max}}}$ is the maximum voltage step, $K_{\mathrm{tr}}$ is the gain applied to the error between the output power, $P$, and the specified power reference, $P_{\mathrm{ref}}$. The classification between the steady-state and the transient modes is done via an auxiliary variable $\sigma$, by which $\sigma = 1$ corresponds to the steady-state mode. The mode classification follows the strategy introduced in \cite{tafti2018adaptive}.




\subsection{Decoupling Irradiance From $\Delta P$}

The P\&O strategy decides what the next voltage step should be based on $\Delta P$ and $\Delta V$ measurements between iterations. However, the performance of this strategy is susceptible to irradiance changes. That is because the output power is affected by both the voltage change ($\Delta V_{pv,n}$) and the irradiance change ($\Delta G_{\mathrm{n}}$) between iterations. Consequently, $\Delta G_{\mathrm{n}}$ will introduce an error when calculating the slope of the P-V curve. To address this issue, we introduce an irradiance decoupling method to remove the impact of $\Delta G$ from the measured $\Delta P$.

When the PV system operates on the left side of the MPP, $I_{pv}$ increases almost linearly with respect to $G$. This can be revealed by the following derivations. First, from (\ref{eq:Iph}), calculate the partial derivative of the photocurrent, $I_\mathrm{ph}$, with respect to $G$ by (\ref{eq:partial_Iph}).
\begin{equation}
    \frac{\partial I_{\mathrm{ph}}}{\partial G} = (1+R_{\mathrm{s0}}/R_{\mathrm{sh0}})I_{\mathrm{sc0}}[1+\alpha_{\mathrm{Isc}}T_{0}(\lambda T-1)]
    \label{eq:partial_Iph}
\end{equation}
Thus, $\frac{\partial I_{\mathrm{ph}}}{\partial G}$ can be considered as a constant assuming that $G$ changes much faster than $T$. 

 Second, we define a relation between the photocurrent ($I_{\mathrm{ph}}$) and the PV output current ($I_{\mathrm{pv}}$) as the currents ratio $K_{\mathrm{ph}}$, which is given by (\ref{PVeq}) and (\ref{eq:Iph}). 
  \begin{equation}
    K_{\mathrm{ph}} = \frac{I_{\mathrm{pv}}}{I_{\mathrm{ph}}}
    \label{eq:Kph}
\end{equation}
 
\begin{figure}[htb]
	\centerline{\includegraphics[width=0.5\textwidth]{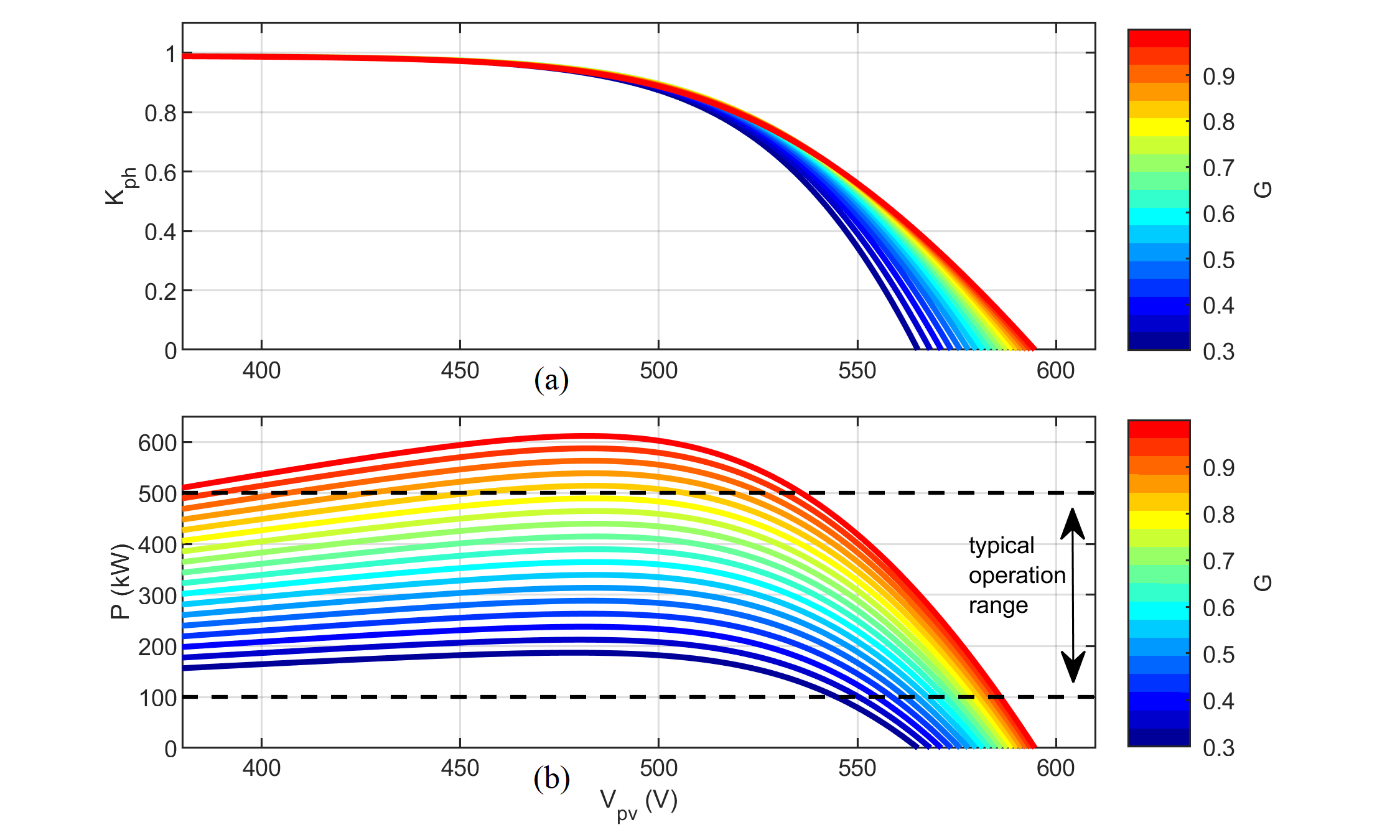}}
	\caption{(a) Relation between $K_{\mathrm{ph}}$ and voltage for irradiance curves in steps of 40 $W/m^2$ (0.04 p.u.). (b) Power-voltage curves of a 500 kVA PV system under different irradiance levels.}
	\label{Kph_rightside}
\end{figure}

Figure \ref{Kph_rightside} shows the $K_{\mathrm{ph}}$ of a 500 kVA PV system for a wide range of voltage and irradiance values. According to the figure, for the same $V_{\mathrm{pv}}$ value, the ratio between $I_{\mathrm{pv}}$ and $I_{\mathrm{ph}}$ is nearly constant with respect to G when the system operates on the left side of the MPP (located around 480V). Thus, as $I_{\mathrm{ph}}=I_{\mathrm{pv}}/K_{\mathrm{ph}}$, based on (\ref{eq:partial_Iph}), we can show that the relationship between $I_{\mathrm{pv}}$ and $G$ is approximately linear.

However, once the operation point passes the MPP (i.e., the PV system operates on the right side of the P-V curve), $I_{\mathrm{pv}}$ no longer linearly increases with respect to $G$. This can be observed by noticing that at same $V_{\mathrm{pv}}$ values, $K_{\mathrm{ph}}$ changes with respect to G (remember that $K_{\mathrm{ph}}$ is the ratio between $I_{\mathrm{pv}}$ and $I_{\mathrm{ph}}$, and that $I_{\mathrm{ph}}$ is always linearly proportional to G). 

To address this issue, we first assume that if irradiance perturbations ($\partial G$) are small (e.g., less than 40 $W/m^{2}$ per FPPT iteration), $K_{\mathrm{ph}}$ is constant at a given $V_{\mathrm{pv}}$ value.
Note that for a small PV system with tens of panels, the irradiance change caused by a passing cloud can be \SI{250}{\watt}/m$^2$/s or more \cite{marion2014new}. However, for a large solar farm with hundreds of panels, the irradiance change will be averaged over all panels, making the irradiance perturbations caused by passing clouds much slower than those in a small PV system \cite{marcos2011irradiance}. Therefore, it is reasonable to assume that $K_{\mathrm{ph}}$ remains constant between FPPT iterations for larger PV systems. 

To decouple the impact of irradiance changes from the PV power output, at iteration $n$, we will first calculate how much the photocurrent from the previous iteration would be if subject to the new irradiance $G_{\mathrm{n}}$ using (\ref{eq:partial_Iph}). Then, given an irradiance change of $\Delta G_{\mathrm{n}} = G_{\mathrm{n}}-G_{n\text{-}1}$, $\Delta I_{\mathrm{pv}}$ can be calculated as


\begin{equation}
    \left.\Delta I_{\mathrm{pv}}\right\rvert_{\Delta G_{\mathrm{n}}} \approx K_{\mathrm{ph},n\text{-}1}\Delta G_{\mathrm{n}} \bigg(\frac{\partial I_{\mathrm{ph}}}{\partial G}\bigg)
    \label{dIdG}
\end{equation}
Note that it is crucial that $K_{\mathrm{ph}}$ from the previous iteration is used to exclude the impacts of $\Delta V_{\mathrm{n}}$. 



When using $V_{\mathrm{pv}}$ and $I_{\mathrm{pv}}$ measurements to compute $\Delta P_{\mathrm{n}}$, the impact of $\Delta G$ is included. Thus, to calculate $\Delta P_{\mathrm{n}}$ without including the impact of irradiance changes,  $I_{\mathrm{pv}}$ needs to be adjusted by $\Delta G$ using (\ref{dIdG}), so we have


\begin{equation}
    \Delta P_{\mathrm{n}} = V_{\mathrm{pv},\mathrm{n}}\Big(I_{\mathrm{pv},\mathrm{n}} - \left.\Delta I_{\mathrm{pv}}\right\rvert_{\Delta G_{\mathrm{n}}}\Big) - V_{\mathrm{pv},\mathrm{n}\text{-}1}I_{\mathrm{pv},\mathrm{n}\text{-}1}
    \label{dPdG}
\end{equation}


For example, a 500 kVA PV system with a 5 Hz FPPT is following a power setpoint of 200 kW. At iteration $n$, $G_{\mathrm{n}}$ is calculated as 0.5 p.u. from (\ref{eq:G}), the available PV power (i.e., $P_{\mathrm{mpp}}$) is 307.8 kW, $V_{pv}$ is 546.6 V, $\tfrac{\partial I_{\mathrm{ph}}}{\partial G}$ is 1359.0 AW$^{\text{-}1}$m$^{2}$ calculated from (\ref{eq:partial_Iph}), and $K_{\mathrm{ph},\mathrm{n}}$ is 0.5386 calculated from (\ref{eq:Kph}). Then, at iteration $n+1$, 
the irradiance increases at a rate of 200 Wm$^{\text{-}2}$s$^{\text{-}1}$ (extreme case for a large PV array) so that 
$G_{\mathrm{n}+1}=0.54$ p.u.. Then, 
calculated by (\ref{dIdG}), $\Delta I_{\mathrm{pv}}|_{\Delta G_{\mathrm{n}}}=29.27$ A. Therefore, the change of power due to irradiance change is calculated by (\ref{dPdG}) as $\Delta P_{\mathrm{n}}=16.0$ kW. 

Note that the proposed irradiance decoupling method performs better for slower irradiance variations or faster FPPT frequency. Thus,
to meet the performance requirement, the minimum FPPT frequency should be designed considering the maximum expected $\Delta G$ of the PV system. More details regarding the estimation of the maximum expected $\Delta G$ based on the size of the PV array can be 
found in \cite{marcos2011irradiance}.

\subsection{Rapid Setpoint Tracking}

\begin{figure}[htb]
	\centerline{\includegraphics[width=0.5\textwidth]{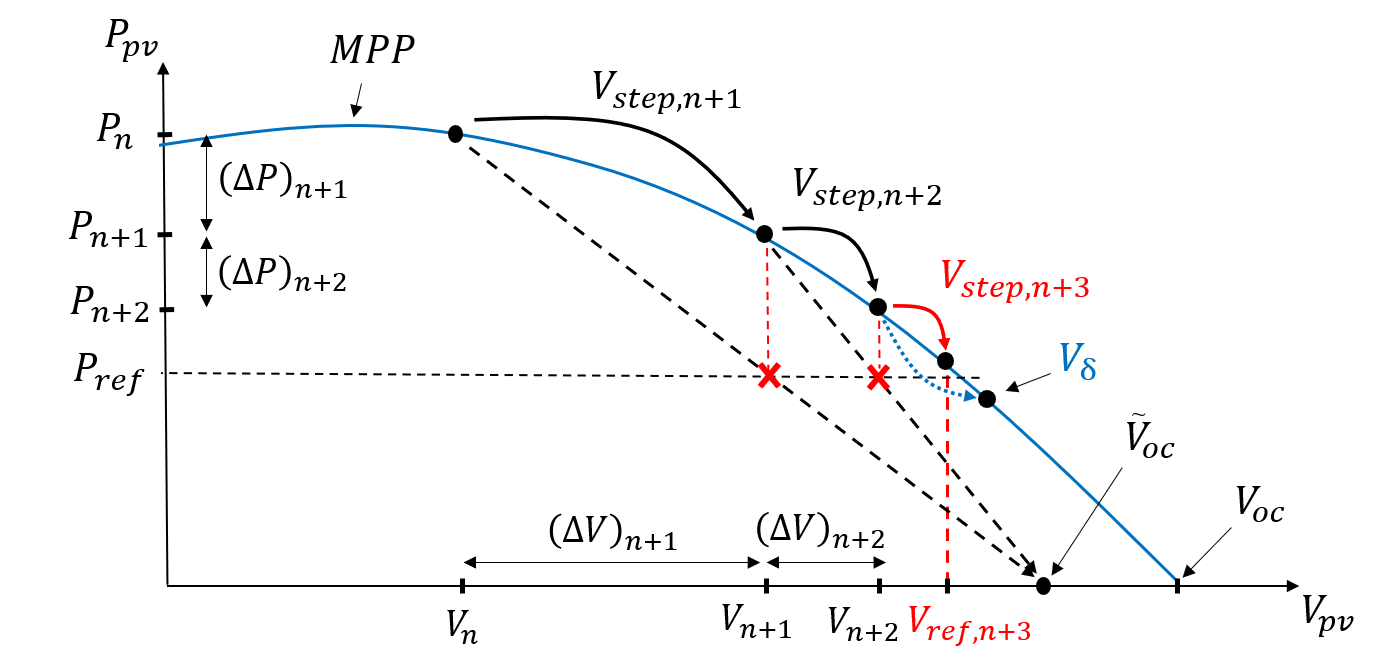}}
	\caption{Sequence of steps of the RST algorithm. Note that slope at $V_{\delta}$ is used for calculating $V_{\mathrm{step},\mathrm{n}+3}$. The figure is not to scale.}
	\label{FFR2}
\end{figure}

In this subsection, we present the Rapid Setpoint Tracking (RST) method for achieving the PV system's voltage reference while maintaining power reserves. This is a novel convergence strategy that can reach the power setpoint given to the PV system within three iterations of the setpoint tracking algorithm, as shown in Fig. \ref{FFR2}. 

At the beginning of each iteration, the open-circuit voltage of the PV system, $\tilde{V}_{\mathrm{oc}}$, is estimated using the equation introduced in \cite{saloux2011explicit} as 

\begin{equation}
    \tilde{V}_{\mathrm{oc}} = ka \; \mathrm{ln}\bigg(1 + \frac{I_{\mathrm{ph}}}{I_{\mathrm{s}}}\bigg)
    \label{kov}
\end{equation}

Note that here we introduce a scaling factor, $k$, for enhancing the robustness of the algorithm and reducing the power overshoots.  This is because if $\tilde{V}_{\mathrm{oc}}$ is higher than the actual open-circuit voltage of the PV system, $V_{\mathrm{oc}}$, a large overshoot may occur. In this paper, $k$ is set at 99\%.

\begin{figure}[htb]
	\centerline{\includegraphics[width=0.5\textwidth]{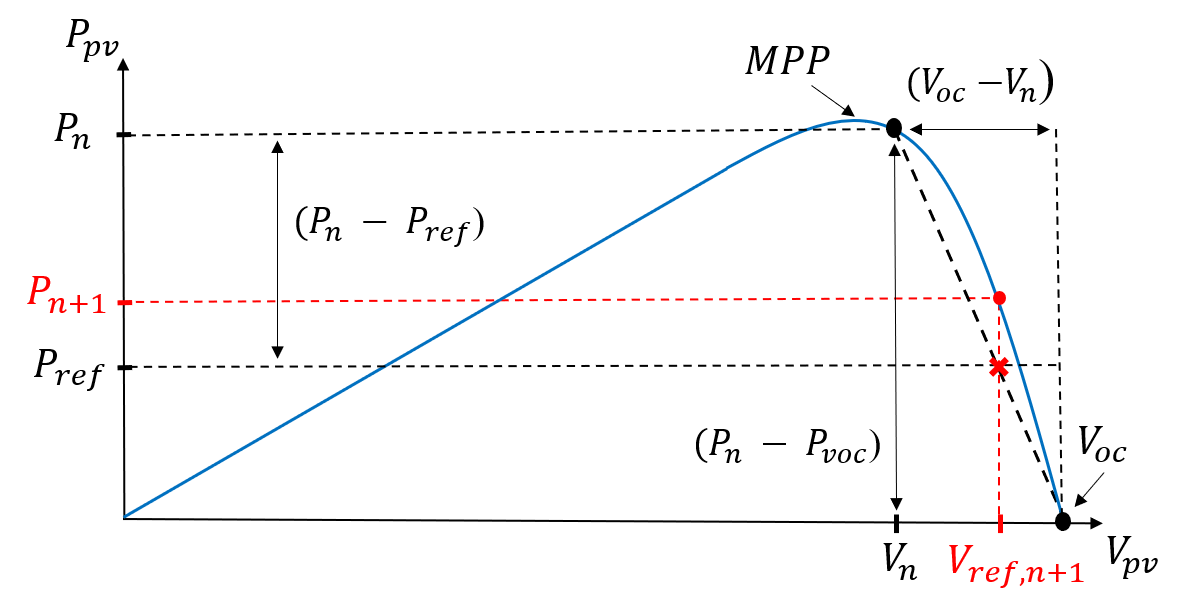}}
	\caption{Calculation of $V_\mathrm{ref}$ in the first step of the RST algorithm.}
	\label{FFR1}
\end{figure}

Then, as illustrated in Fig. \ref
{FFR1}, the voltage reference setpoint, $V_{\mathrm{ref}}$, is calculated by (\ref{first_step}), where $V_{\mathrm{n}}$ and $P_{\mathrm{n}}$ are measured PV voltage and power values at step $n$. 
\begin{equation}
    V_{\mathrm{ref},\mathrm{n}+1} = V_{\mathrm{n}} + \frac{(\tilde{V}_{\mathrm{oc}}-V_{\mathrm{n}})(P_{\mathrm{n}}-P_{\mathrm{ref}})}{P_{\mathrm{n}}}
    \label{first_step}
\end{equation}

Using $V_{\mathrm{ref}}$ calculated by (\ref{first_step}), we can achieve fast convergence towards $P_{\mathrm{ref}}$ in the first two steps. However, because $\tilde{V}_{\mathrm{oc}}$ will include estimation error, and because the use of $k$ for improving robustness may introduce additional error, the convergence slows down at the third step. Therefore, a different strategy for calculating $V_\mathrm{{ref, \mathrm{n}+1}}$ is needed in the third step to reduce the impact of estimation error and quickly converge to $P_{\mathrm{ref}}$. Thus, at the beginning of the third step, an artificial step ($V_{\delta}$) is first calculated using the $\tfrac{\Delta P}{\Delta V}$ slope measured in the previous iteration, i.e., iteration $n+2$ of the fast-convergence method, as

\begin{equation}
V_{\delta} = \frac{(V_{\mathrm{n}+2}-V_{\mathrm{n}+1})(P_{ref}-P_{\mathrm{n}+2})}{(P_{\mathrm{n}+2}-P_{\mathrm{n}+1})}
\end{equation}

Subsequently, to account for the non-linearity on the right side of the P-V curve, we use the first-order Taylor series approximation to estimate the increase in the slope of the P-V curve when the $V_{\delta}$ step is applied by

\begin{equation}
    \bigg(\dfrac{\Delta P}{\Delta V}\bigg)_{\delta} = \bigg(\dfrac{\Delta P}{\Delta V}\bigg)_{\mathrm{n}+2} + \Bigg[\dfrac{\big(\frac{\Delta P}{\Delta V}\big)_{\mathrm{n}+2} - \big(\frac{\Delta P}{\Delta V}\big)_{\mathrm{n}+1}}{V_{\mathrm{n}+2}-V_{\mathrm{n}+1}}\Bigg]V_{\delta}
    \label{eq:projection}
\end{equation}

Then, the actual voltage step for iteration three, $V_{step,n+3}$, can be calculated by (\ref{eq:step3}), and the voltage reference is updated by (\ref{eq:Vref3}).

\begin{equation}
    V_{\mathrm{step},\mathrm{n}+3} = (P_{\mathrm{n}+2}-P_{\mathrm{ref}})\Bigl|\Big(\tfrac{\Delta V}{\Delta P}\Big)_{\delta}\Bigr|
    \label{eq:step3}
\end{equation}

\begin{equation}
    V_{\mathrm{ref},\mathrm{n}+3} = V_{\mathrm{n}+2} +    V_{\mathrm{step},\mathrm{n}+3}
    \label{eq:Vref3}
\end{equation}


After the third iteration, the RST algorithm will be terminated, and the tracking mechanism returns to the P\&O-based technique. The RST algorithm can be terminated before the three-step convergence ends under the following two conditions:

\begin{enumerate}
    \item At any step, if $V_{\mathrm{ref}}\leq V_{\mathrm{mpp}}$ (which may happen if the power setpoint is larger than the available PV power), the new voltage reference is set as $V_{\mathrm{mpp}}$. This condition is specifically set for a system operating on the right-side of the MPP. Note that for systems designed to operate on the left side, the condition must be changed to $V_{\mathrm{ref}} \geq V_{\mathrm{mpp}}$.
    \item The operating mode is classified as steady-state (i.e., $\sigma = 1$).
\end{enumerate}

Moreover, if at any iteration $V_{\mathrm{pv}}$ is greater than $\tilde{V_{\mathrm{oc}}}$ because of the safe margin added, which could happen under high irradiance and very high levels of power curtailment (under 0.1 p.u.), then $\tilde{V}_{\mathrm{oc}}$ is set as $1.005\times V_{\mathrm{pv}}$. This setup will ensure $\tilde{V}_{\mathrm{oc}}>V_{\mathrm{pv}}$ in order to prevent the algorithm from mistakenly stepping towards higher voltage values and ceasing the power generation of the PV array.

Note that the system should not always be in RST. This is because a good measurement window is needed for the estimator to converge based solely on voltage and current measurements. If all measurements are concentrated within a small operating region, the curve-length coverage will be insufficient, leading to a degradation in convergence performance. Therefore, the algorithm is set to operate under the P\&O mode when fast convergence is not needed. By leveraging the oscillatory performance of P\&O, we demonstrate that it is possible to maintain a good measurement window to achieve sufficient convergence.

\section{Real-Time Simulation Results}

To validate the proposed algorithm, a testbed of a 500 kVA PV system is set up in an OPAL-RT real-time simulation platform. One-second resolution data collected from a 1.04 MW solar farm by EPRI is used \cite{EPRIdata}. This publicly available data set provides 9 days-long, high-resolution measurement data collected from PV systems ranging from 0.18 to 1040 kW. The simulation timestep is \SI{50}{\micro s}. To account for measurement noise from the dc sensors, Gaussian noises are added to $V_{\mathrm{pv}}$, $I_{\mathrm{pv}}$, and $V_{\mathrm{dc}}$, following the SNR ratio from a TI AMC1303 sensor \cite{TIsensor}. Table \ref{parameters} displays the simulation parameters.

\renewcommand{\arraystretch}{1.05}
\begin{table}[htb]
	\caption{Simulation Parameters}
	\begin{center}
		\begin{tabular}{|>{\columncolor[gray]{0.85}} c|c|c|}
			\hline
			
			&Power&\SI{612}{\kilo\watt}\\\cline{2-3}
			&Module&CS6P-250P\\\cline{2-3}
			&Size (parallel$\times$series) &153 $\times$ 16\\\cline{2-3}
			\multirow{-4}{*}{PV Array}&V$_{\mathrm{mpp}}$, I$_{\mathrm{mpp}}$& \SI{481.6}{\volt}, \SI{1270}{A}\\
			\hline\hline
			&Power, Frequency&\SI{500}{\kilo VA}, \SI{60}{\hertz} \\\cline{2-3}
			&L$_{\mathrm{f}}$, r$_{\mathrm{L}}$&\SI{100}{\micro\henry}, \SI{3}{\milli\ohm}\\\cline{2-3}
			&C$_{\mathrm{dc}}$& \SI{5000}{\micro\farad}\\\cline{2-3}
			&PI (v$_{\mathrm{dc}}$)& K$_{\mathrm{p}}$ = 1, K$_\mathrm{{i}}$ = 250\\\cline{2-3}
			\multirow{-5}{*}{Inverter}&PI (i$_{\mathrm{d}}$, i$_{\mathrm{q}}$)& K$_{\mathrm{p}}$ = 0.7, K$_{\mathrm{i}}$ = 50\\
			\hline\hline
			&$\Delta G_{\mathrm{max}}$ & \SI{200}{\watt}/m$^{2}$/s \\\cline{2-3}
			&$\Delta T_{\mathrm{max}}$ & \SI{3}{\degree C}/min \\\cline{2-3}
			& Measurement window & 100 samples \\\cline{2-3}
			&$a$ ($\eta$ gain) & 3 \\\cline{2-3}
			&Damping $\eta$ range& [$10^{{\text-}6}, 10^{\text{-}3}$]\\\cline{2-3}
			&Sampling freq. ($f_{\mathrm{s}} = 1/$T$_{\mathrm{s}}$)& \SI{20}{\hertz}\\\cline{2-3}
			\multirow{-7}{*}{\shortstack{Maximum\\ Power\\ Point\\ Estimation}}&\shortstack{LM period ($T_{\mathrm{LM}}$)}& \SI{5}{\second}\\
			\hline\hline
			&Frequency ($f_{\mathrm{step}} = 1/$T$_{\mathrm{step}}$)& \SI{4}{\hertz} \\\cline{2-3}
			&$V_{\mathrm{step}\text{-}\mathrm{min}}$ / $V_{\mathrm{step}\text{-}\mathrm{b}}$ & 0.75 / 2 V\\\cline{2-3}
			& $K_{\mathrm{tr}}$  & 0.002\\\cline{2-3}
			& Threshold $\Delta P_{\mathrm{ref,th}}$ & \SI{50}{k\watt/s} \\\cline{2-3}
			&Threshold dp$_{\mathrm{th}}$ & \SI{15}{k\watt}\\\cline{2-3}
			\multirow{-6}{*}{\shortstack{Flexible\\ Power\\ Point\\ Tracking}}&Threshold $dp/dv$& \SI{667}{\watt / \volt}\\
			
			\hline\hline
			&Power, Frequency & 3.125 MVA, \SI{60}{\hertz} \\\cline{2-3}
			&Voltage (line-line)& 2400 V (RMS)\\\cline{2-3}
			&Droop & 5\%\\\cline{2-3}
			&Inertia (H) & 1.07 p.u.\\\cline{2-3}
			& Governor ($T_{\mathrm{g}}$)  & 0.2\\\cline{2-3}		\multirow{-6}{*}{Diesel Generator}&Turbine ($T_{\mathrm{t}}$) & 0.35\\
			\hline
		\end{tabular}
	\end{center}
	\label{parameters}
\end{table}
		
\subsection{Performance of the MPPE}

\begin{figure}[htb]
	\centerline{\includegraphics[width=0.45\textwidth]{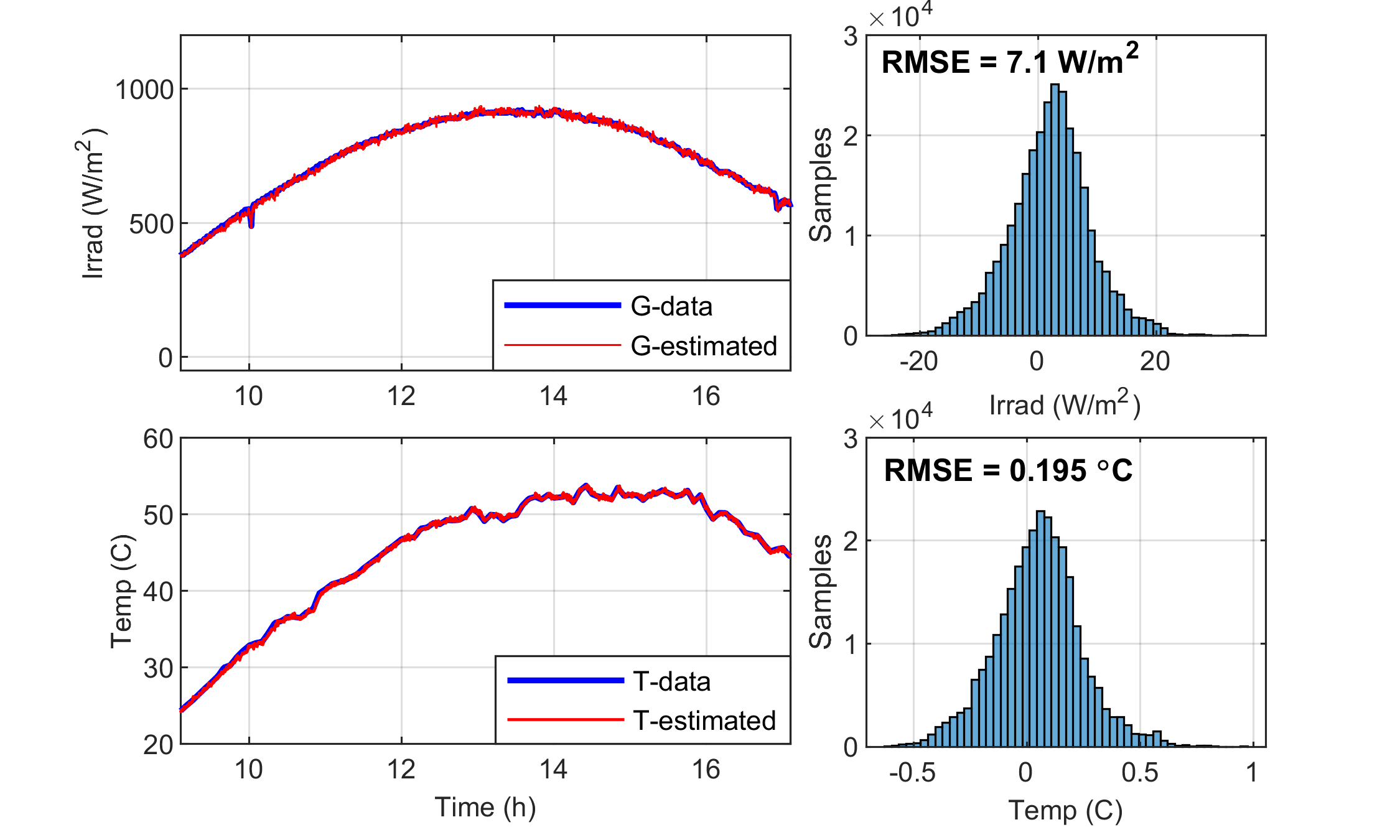}}
	\caption{Performance of the MPPE algorithm in a sunny day (Sampling rate: 10 Hz).}
	\label{estimation_sunny}
\end{figure}

\begin{figure}[htb]
	\centerline{\includegraphics[width=0.45\textwidth]{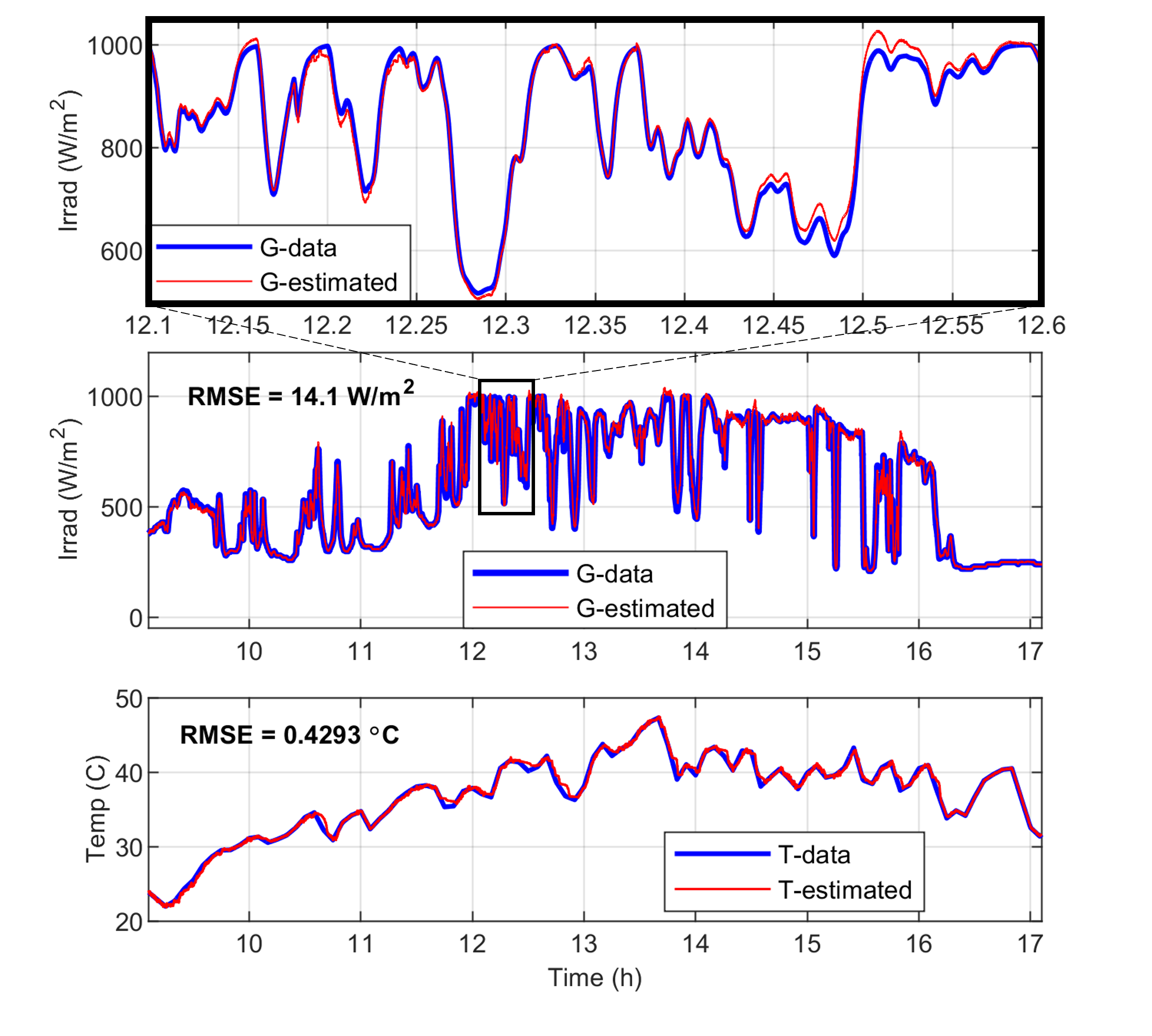}}
	\caption{Performance of the MPPE algorithm in a cloudy day with estimated irradiance threshold set to 1100 W/m$^2$.}
	\label{estimation_rainy}
\end{figure}

\begin{figure}[htb]
	\centerline{\includegraphics[width=0.45\textwidth]{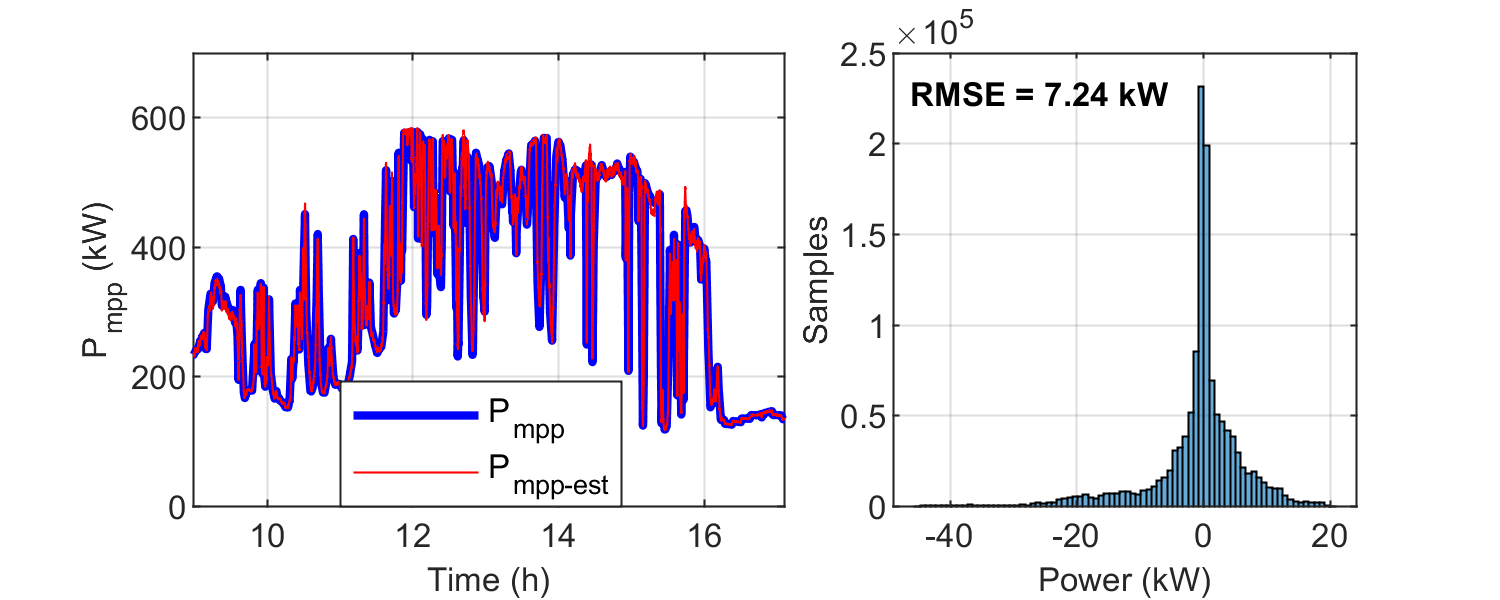}}
	\caption{Comparison between the actual MPP and the estimated MPP for a cloudy day. (Sampling rate: 40 Hz)}
	\label{Pmpp}
\end{figure}

To evaluate the performance of the MPPE, the PV system is set to operate for 10 hours with a fixed 200 kW headroom under two different types of environmental conditions: sunny and cloudy days. Figures \ref{estimation_sunny} and \ref{estimation_rainy} present the performance of the MPPE algorithm from each case. As expected, the estimated irradiance and temperature RMSE is greater in cloudy days due to faster environmental changing conditions, yet, the MPPE can successfully maintain an adequate estimation and not lose convergence throughout this extreme case. A comparison between the actual MPP and the estimated MPP is displayed in Fig. \ref{Pmpp}.

\begin{figure}[htb]
	\centerline{\includegraphics[width=0.5\textwidth]{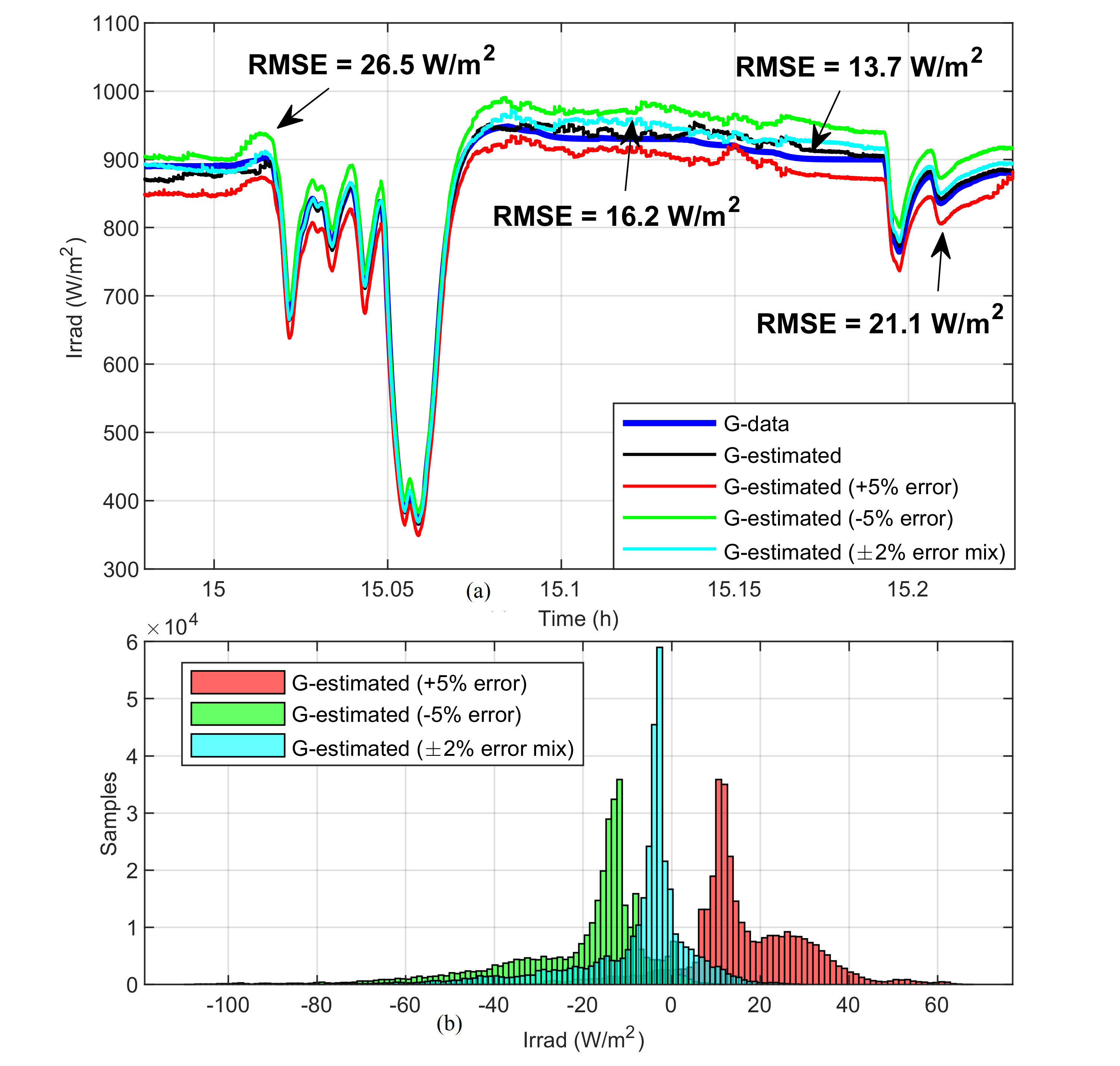}}
	\caption{Performance of the MPPE algorithm in a cloudy day considering PV model parameter errors (Sampling rate: 10 Hz).}
	\label{paramerror}
\end{figure} 

In practice, the base values of the five parameters of the single-diode PV model ($a_{0}$, $R_{\mathrm{s0}}$, $R_{\mathrm{sh0}}$, $I_{\mathrm{ph0}}$, and $I_{\mathrm{s0}}$), corresponding to STC, are also calculated based on measurements of short-circuit current ($I_{\mathrm{sc0}}$), open-circuit voltage ($V_{\mathrm{oc0}}$), voltage at MPP ($V_{\mathrm{mp0}}$), and current at MPP ($I_{\mathrm{mp0}}$) instead of obtaining ($I_{\mathrm{sc0}}$, $V_{\mathrm{oc0}}$, $V_{\mathrm{mp0}}$ and $I_{\mathrm{mp0}}$) from datasheet information. Therefore, we first add $\pm 5\%$ error to each PV parameter base value and redo the test for the cloudy days to simulate two worst case scenarios: significantly over-estimate and under-estimate. Then, we run a case assuming that the base parameters have been optimized so that the estimation errors are within $\pm2$\%.

\begin{figure}[htb]
	\centerline{\includegraphics[width=0.5\textwidth]{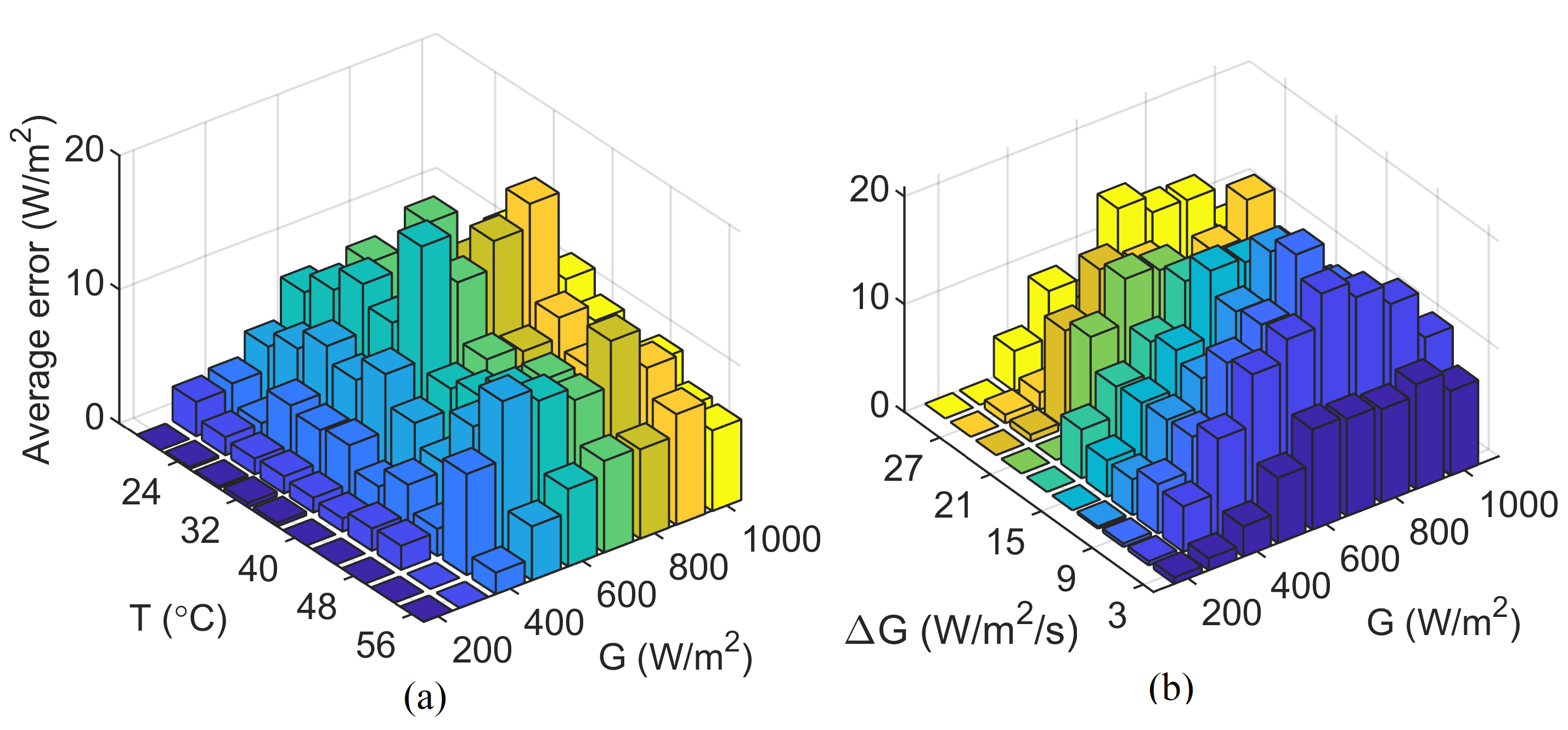}}
	\caption{Irradiance estimation error with respect to: (a) temperature and irradiance, (b) irradiance changes ($\Delta G$) and irradiance.}
	\label{fig:error_data}
\end{figure}

The results of the irradiance estimation under parameter errors are shown in Fig. \ref{paramerror}(a). Clearly, the worst case scenarios are when we over- or under- estimate the PV model parameter by 5\% so that the estimated irradiance presents an offset (Fig. \ref{paramerror}(b)). In such cases, the overall estimation RMSE is higher. However, if the PV model base parameter errors are maintained within $\pm$2\%, the RMSE (16.2 W/m$^2$) is very close to the ideal parameter case (13.7 W/m$^2$). Therefore, it is important to perform PV model parameter correction periodically when applying the proposed algorithm. In addition, the irradiance estimation error with respect to temperature and irradiance conditions are analyzed using results from six different daily irradiance patterns so that sufficient data points can be collected for different temperature and irradiance levels. As shown in Fig. \ref{fig:error_data}, the estimation error is higher at higher irradiance values, and the irradiance change is above 3W/m$2$/s. This is because those time periods with low irradiance changes typically represent overcast and sunny weather conditions, when forecasting results are consistent. The results also show that the MPPE estimation error does not have obvious correlation with temperature.

\subsection{Irradiance Decouple Method}

\begin{figure}[htb]
	\centerline{\includegraphics[width=0.5\textwidth]{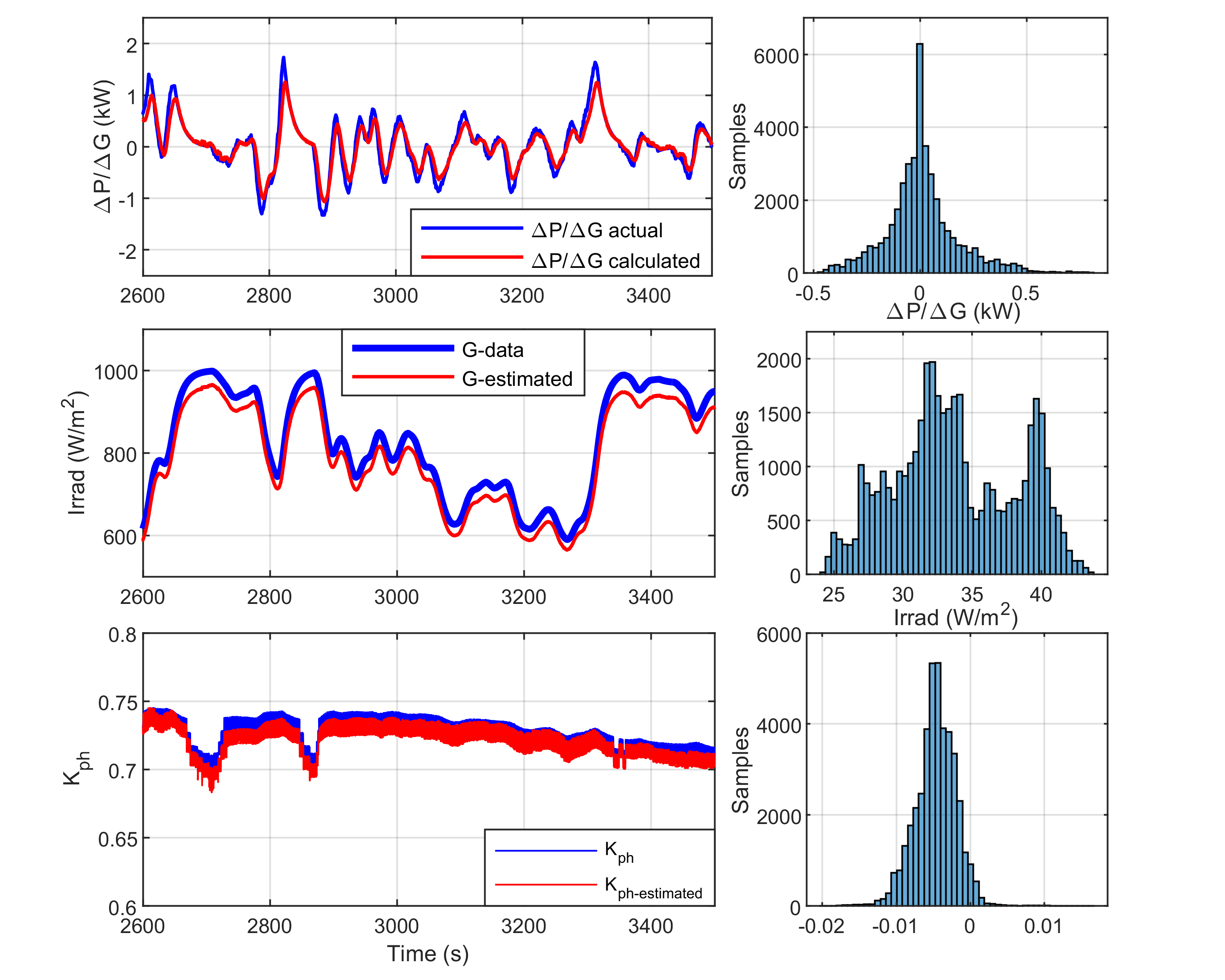}}
	\caption{Performance of irradiance decouple method under 5\% parameter error. (Sampling rate: 40 Hz)}
	\label{Gdecouple}
\end{figure}

Next, we evaluate the accuracy of the irradiance decoupling method presented in Section III-C. The decoupling method is used to estimate how much of the measured $\Delta P$ at a given iteration corresponds to irradiance changes ($\Delta G$), which can be calculated by $V_{\mathrm{pv}}\hspace{-0.08cm}\left.\Delta I_{\mathrm{pv}}\right\rvert_{\Delta G}$ using (\ref{dPdG}). The FPPT frequency is set as 4 Hz. Figure \ref{Gdecouple} compares the calculated $\Delta P$ caused by $\Delta G$ in each iteration with its actual value while the system operates under $5\%$ parameter errors. Notice that the errors from the irradiance estimation do not directly affect the decoupling technique. That is because the decoupling method accounts only for the change in irradiance. The accuracy of the results confirms the assumption that $K_{\mathrm{ph}}$ can be considered constant in the subsequent step due to the small $\Delta G$ between iterations. It must be noticed that a low-pass filter is applied to the estimated irradiance and temperature values used by the irradiance decouple technique. Because the temperature update in the LM iteration is much slower than that in the irradiance decouple method, adding the filter can effectively prevent sudden changes caused by step changes in the LM iterations from propagating to the irradiance decoupling technique.  Therefore, the impact of the last temperature update will not lead to changes in temperature between the last two iterations in the irradiance decouple technique. For the estimated temperature, a filter with bandwidth 1-10 times slower than the frequency of the LM updates provided good results. A good value for the filter utilized for the estimated irradiance can be found in \cite{marcos2011irradiance}.

\subsection{Performance of the Rapid Setpoint Tracking algorithm}

\begin{figure}[htb]
	\centerline{\includegraphics[width=0.5\textwidth]{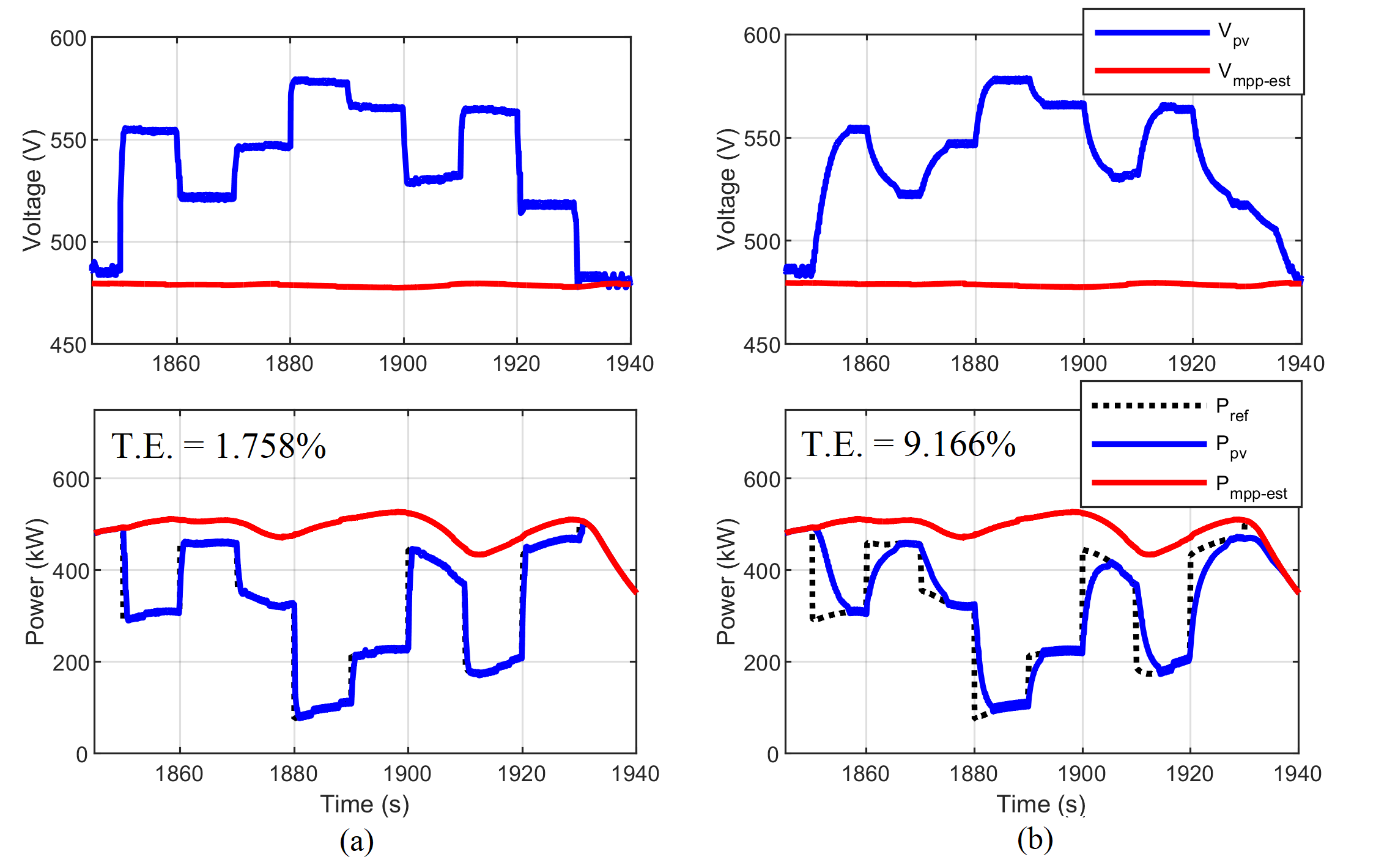}}
	\caption{Comparison of the power setpoint tracking performance under irradiance intermittency between (a) the proposed RST method and (b) the adaptive FPPT (state-of-the-art).}
	\label{ptrack}
\end{figure}

To analyze the performance of the RST algorithm  when providing regulation services, we let the PV system follow a sequence of power setpoint changes while undergoing irradiance variations (i.e. MPP is varying). The PV system has a power setpoint tracking FPPT with identical settings from Table \ref{parameters}. Two cases are modeled: one with the proposed RST and the other with the adaptive FPPT (the state-of-the-art). The tracking errors (T.E.) of the two algorithms are calculated using the method presented in \cite{paduani2020maximum} and the power setpoint tracking results are shown in the bottom two plots in Fig. \ref{ptrack}. 
The simulation results demonstrate that RST can closely track power setpoint changes while the adaptive FPPT shows significant delays. This is because the later calculates the voltage step in proportion to the error between its output power and the given power setpoint. Therefore, when the error decreases, the step size also decreases, slowing down the convergence towards the end of the process. Using the 3-step convergence strategy, RST achieved a much faster convergence speed.

\subsection{PV-based Fast Frequency Response Service}
\begin{figure}[!t]
	\centerline{\includegraphics[width=0.5\textwidth]{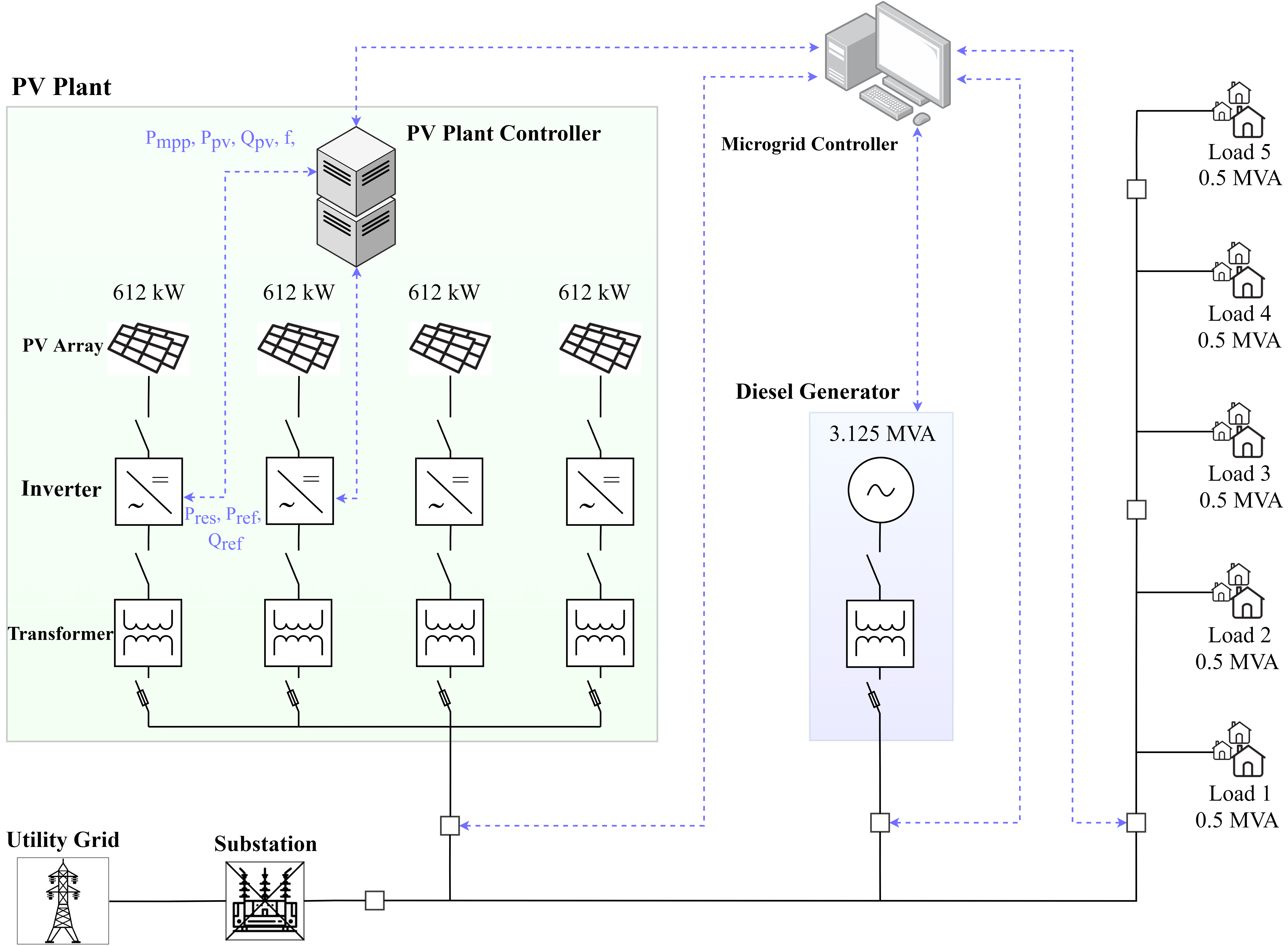}}
	\caption{Configuration of a feeder-level microgrid.}
	\label{fig:microgrid}
\end{figure}

One of the main applications of the proposed unified MPPE and FPPT control is to enable PV plants to provide high-quality, fast frequency response (FFR) services.  To demonstrate the performance of the PV-based  FFR, the black start process of an islanded feeder-level microgrid is modeled on the  OPAL-RT eMEGASIM platform. The microgrid, displayed in Fig. \ref{fig:microgrid}, is powered by a utility-scale PV plant with 4 identical 612-kW PV arrays and a 3.125-MVA diesel generator. The system parameters are listed in Table \ref{parameters}. The feeder is divided into five load groups, each supplying 0.5 MVA loads. The cold-load pickup effect of each load group is modeled using the delayed exponential method introduced in \cite{edstrom2012modeling}, which can double the load of each load group to 1 MVA during the initial transients.  In this case, we set the FPPT frequency to be 10 Hz. Note that typically, to achieve satisfactory performance, the FPPT/MPPT bandwidth is set between 1 to 10 Hz \cite{sangwongwanich2018analysis}. Moreover, note that the PV plant controller sends individual power commands (P$_\mathrm{res}$ or P$_\mathrm{ref}$, and Q$_\mathrm{ref}$) to each inverter of the plant.

\begin{figure}[!t]
	\centerline{\includegraphics[width=0.5\textwidth]{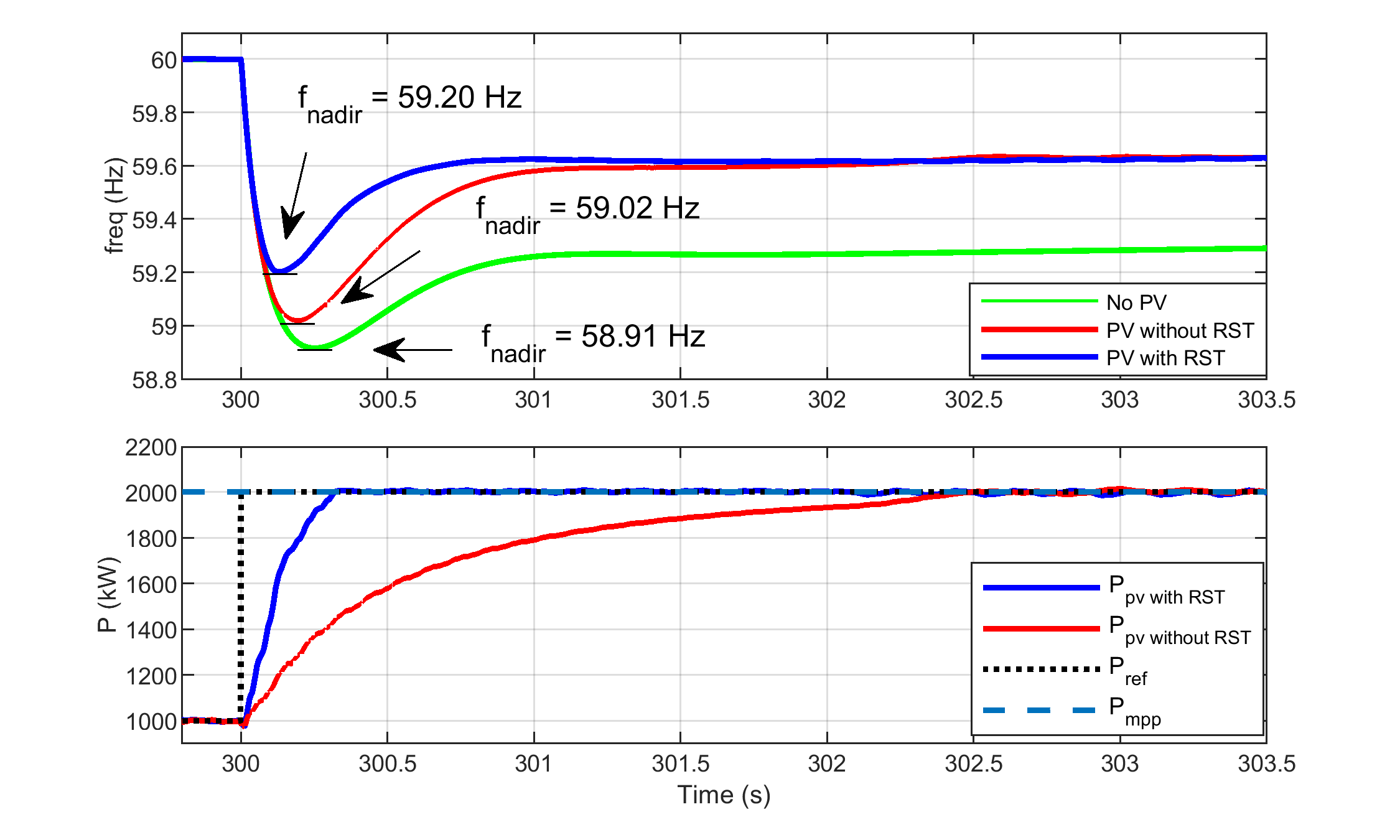}}
	\vspace{-10pt}
	\caption{Frequency and voltage plots during microgrid cold-load pickup.}
	\label{fig:FFR}
\end{figure}

After the first two load groups have been picked up, the PV plant power setpoint is set to be 1 MW to maintain 1 MW of power reserves. Then, the microgrid controller dispatches the 1-MW PV power reserves to reduce the frequency nadir during the second cold-load pickup procedure, in which load groups 3 and 4 are connected to the microgrid. It is assumed that during the load pick-up coordination, the microgrid controller considers the communication delay, i.e., how long it would take between issuing the commands to the relays and PV plant controller via MODBUS, and their execution time. As shown in Fig. \ref{fig:FFR}, the proposed method outperforms the state-of-the-art method by reducing the frequency nadir during the cold-load pickup.

The IEEE Standard 1547-2018 expects PV inverters to provide frequency ride-through function when the rate of change of frequency (ROCOF) is 2 to 3 Hz per second depending on DER penetration levels. Note that large ROCOF events are uncommon in high-inertia grids but occur more frequently in low-inertia grids dominated by inverter-based resources, as having been seen in the Hawaiian grid \cite{pattabiraman2018impact}. Therefore, in the next analysis we suppose the microgrid from Fig. \ref{fig:microgrid} is connected to a low inertia grid susceptible to large ROCOF events with low frequency nadirs. 

\begin{figure}[!t]
	\centerline{\includegraphics[width=0.5\textwidth]{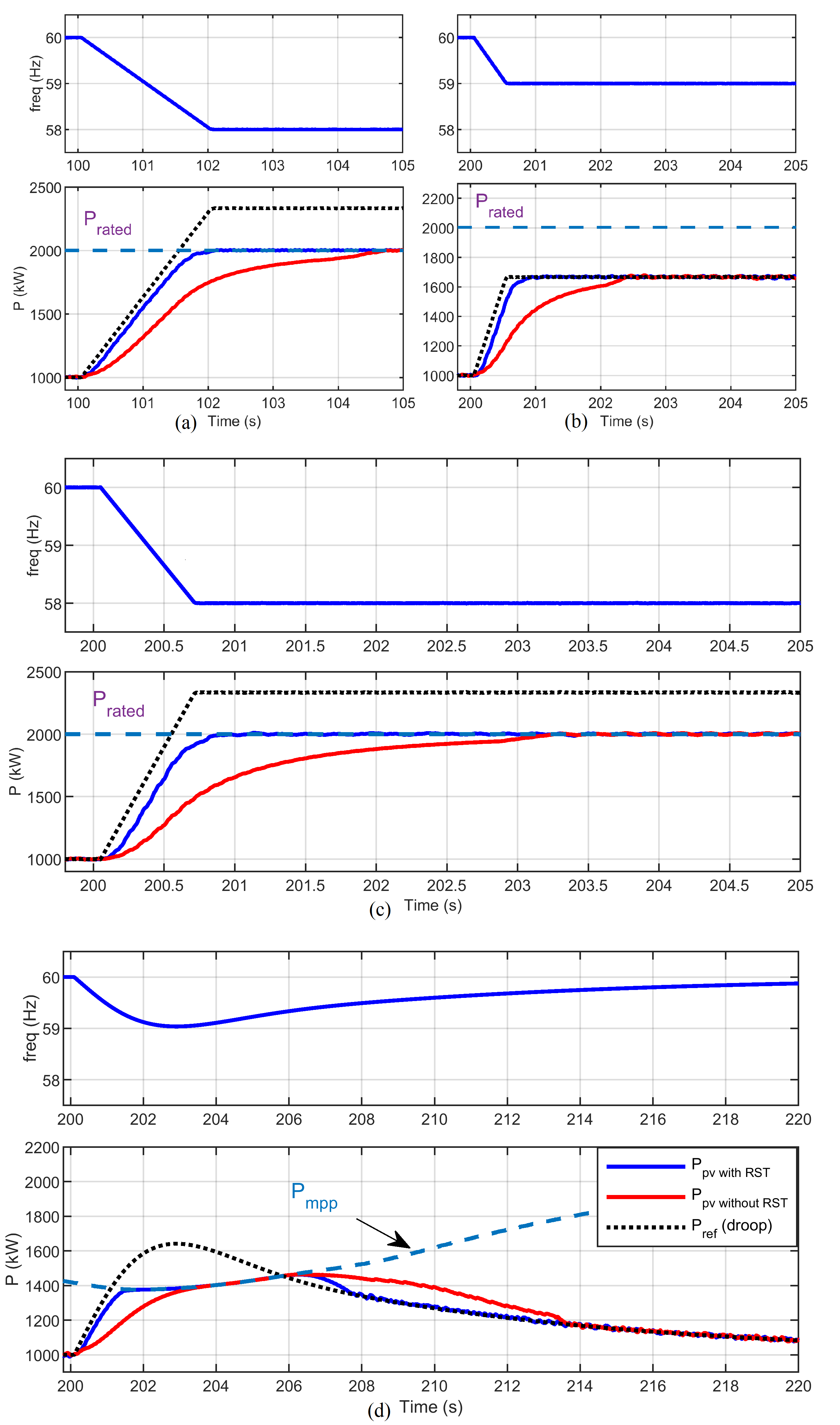}}
	\caption{Frequency-watt droop responses of a PV plant. (a)  Case 1: 1 Hz/s ROCOF and 58 Hz nadir. (b) Case 2: 2 Hz/s ROCOF and 59 Hz nadir. (c) Case 3: 3 Hz/s ROCOF and 58 Hz nadir. (d) Case 4: droop response in a grid disturbance under rapid irradiance changes.}
	\label{fig:rocof1}
\end{figure} 

The frequency-watt droop of each PV inverter is set to 5\%  without a deadband, which corresponds to a PV power setpoint change of 1/3 p.u. per \SI{1}{\hertz} change in a 60 Hz grid. We test four cases: 1) ROCOF of \SI{1}{\hertz/s} and nadir of \SI{58}{\hertz}, 2) ROCOF of \SI{2}{\hertz/s} and nadir of \SI{59}{\hertz}, 3) ROCOF of \SI{3}{\hertz/s} and nadir of \SI{58}{\hertz}, and 4) droop response under rapid irradiance changes. As shown in Fig. \ref{fig:rocof1}, the PV farm is able to provide faster power setpoint tracking with the RST function.

The delay observed between $P_{\mathrm{ref}}$ and $P_{\mathrm{pv}}$ in Figs. \ref{fig:rocof1}(a) and \ref{fig:rocof1}(b) is caused by the low FPPT frequency (10 Hz), which cannot be increased further because of the need for decoupled bandwidths in hierarchical control systems \cite{paduani2019small}. Note that the delay between second 206 and 207 in Fig. \ref{fig:rocof1}(c) is caused by the RST activation delay.  This is because the RST function is only activated when the operation mode is classified as transient (Fig. \ref{flowchart}), the threshold of which is determined by the value of $dp_{\mathrm{th}}$, given in Table \ref{parameters}. Once the RST is activated, the PV system will quickly track the droop command during frequency events. This use case demonstrates the importance of the RST function for improving grid resiliency in high solar participation grids. When operated with power reserves, the proposed algorithm allows utility-scale PV farms to follow frequency-watt droop response curves in high ROCOF events. This function is crucial for reducing the frequency nadir in low-inertia power grids.




\section{Conclusion}

Being able to closely follow power setpoints is an essential function for a PV system to provide grid services such as curtailment, spinning/non-spinning reserves, and fast frequency response. Therefore, in this paper, we combine MPPE and FPPT functionalities into a unified fast power setpoint tracking algorithm that significantly improves the performance of each functionality. Simulation results show that the proposed real-time, nonlinear curve-fitting MPPE estimates irradiance and temperature accurately even in the presence of modelling errors and measurement noises without the need for an external ripple. Moreover, the MPPE improves the P\&O performance by decoupling the impact of the irradiance change from the PV output power. Finally, aided by MPPE, the proposed rapid setpoint tracking method can follow power setpoint changes in just three iterations, outperforming the state-of-the-art method. For the utilization of the proposed method in real-life applications, a validation in hardware is necessary.



%

\appendix[]

The computational cost of the LM parameter update process that corresponds to the slowest period of the proposed algorithm ($T\mathrm{_{LM}}$) is summarized in Table \ref{table2}. The number of operations for each LM iteration is a function of the size of the measurement window, $N$. The number of CPU cycles required to execute each operation on a Texas Instrument TMS320C28x Floating Point Unit \cite{instruments2008tms320c28x} is used as a benchmark. The Lambert $W$ and the logarithmic functions are not included because their calculations should be substituted by look-up tables for the implementation in a microprocessor.

\renewcommand{\arraystretch}{1.05}
\begin{table}[htb]
    \caption{Required CPU cycles for one LM iteration.}
    \begin{center}
    \begin{tabular}{|c|c|c|c|}
    
        \hline
        Operation & Cycles & No of Operations & Total Cycles  \\
        \hline
         + / -& 2 & 40 + N$\times$30 & 80 + N$\times$60\\
        \hline
        $\times$ & 24 & 56 + N$\times$51 & 1344 + N$\times$1224\\
        \hline
        $\div$ & 63 & 12 + N$\times$6 & 756 + N$\times$378\\
        \hline
        Max / Min & 7 & 6 + N$\times$3 & 42 + N$\times$21\\
        \hline
        Comparator& 1 & 7 & 7\\
        \hline
        Absolute & 2 & 12 & 24\\
        \hline
        
    \end{tabular}
    \end{center}
    
    \label{table2}
\end{table}




\ifCLASSOPTIONcaptionsoff
  \newpage
\fi
\bibliographystyle{IEEEtran}
\bibliography{powercurtailment_FFR}

\begin{thebibliography}{10}
\providecommand{\url}[1]{#1}
\csname url@samestyle\endcsname
\providecommand{\newblock}{\relax}
\providecommand{\bibinfo}[2]{#2}
\providecommand{\BIBentrySTDinterwordspacing}{\spaceskip=0pt\relax}
\providecommand{\BIBentryALTinterwordstretchfactor}{4}
\providecommand{\BIBentryALTinterwordspacing}{\spaceskip=\fontdimen2\font plus
\BIBentryALTinterwordstretchfactor\fontdimen3\font minus
  \fontdimen4\font\relax}
\providecommand{\BIBforeignlanguage}[2]{{%
\expandafter\ifx\csname l@#1\endcsname\relax
\typeout{** WARNING: IEEEtran.bst: No hyphenation pattern has been}%
\typeout{** loaded for the language `#1'. Using the pattern for}%
\typeout{** the default language instead.}%
\else
\language=\csname l@#1\endcsname
\fi
#2}}
\providecommand{\BIBdecl}{\relax}
\BIBdecl

\bibitem{ieee1547ieee}
I.~S. Association \emph{et~al.}, ``{IEEE Std}. 1547-2018,'' \emph{Standard for
  interconnection and interoperability of distributed energy resources with
  associated electric power systems interfaces}, 2018.

\bibitem{california2016electric}
{California Public Utilities Commission} \emph{et~al.}, ``Electric rule no. 21
  generating facility interconnections,'' 2016.

\bibitem{ishaque2012improved}
K.~Ishaque, Z.~Salam, M.~Amjad, and S.~Mekhilef, ``An improved particle swarm
  optimization ({PSO})--based {MPPT} for {PV} with reduced steady-state
  oscillation,'' \emph{IEEE Transactions on Power Electronics}, vol.~27, no.~8,
  pp. 3627--3638, 2012.

\bibitem{hoke2013active}
A.~Hoke and D.~Maksimovi{\'c}, ``Active power control of photovoltaic power
  systems,'' in \emph{2013 1st IEEE Conference on Technologies for
  Sustainability (SusTech)}.\hskip 1em plus 0.5em minus 0.4em\relax IEEE, 2013,
  pp. 70--77.

\bibitem{hoke2017rapid}
A.~F. Hoke, M.~Shirazi, S.~Chakraborty, E.~Muljadi, and D.~Maksimovic, ``Rapid
  active power control of photovoltaic systems for grid frequency support,''
  \emph{IEEE Journal of Emerging and Selected Topics in Power Electronics},
  vol.~5, no.~3, pp. 1154--1163, 2017.

\bibitem{blanes2012site}
J.~M. Blanes, F.~J. Toledo, S.~Montero, and A.~Garrig{\'o}s, ``In-site
  real-time photovoltaic i--v curves and maximum power point estimator,''
  \emph{IEEE Transactions on Power Electronics}, vol.~28, no.~3, pp.
  1234--1240, 2012.

\bibitem{de2012evaluation}
M.~A.~G. De~Brito, L.~Galotto, L.~P. Sampaio, G.~d.~A. e~Melo, and C.~A.
  Canesin, ``Evaluation of the main mppt techniques for photovoltaic
  applications,'' \emph{IEEE Transactions on industrial electronics}, vol.~60,
  no.~3, pp. 1156--1167, 2012.

\bibitem{soon2014fast}
T.~K. Soon and S.~Mekhilef, ``A fast-converging {MPPT} technique for
  photovoltaic system under fast-varying solar irradiation and load
  resistance,'' \emph{IEEE Transactions on industrial informatics}, vol.~11,
  no.~1, pp. 176--186, 2014.

\bibitem{ahmed2018enhanced}
J.~Ahmed and Z.~Salam, ``An enhanced adaptive {P\&O} {MPPT} for fast and
  efficient tracking under varying environmental conditions,'' \emph{IEEE
  Transactions on Sustainable Energy}, vol.~9, no.~3, pp. 1487--1496, 2018.

\bibitem{sangwongwanich2017benchmarking}
A.~Sangwongwanich, Y.~Yang, F.~Blaabjerg, and H.~Wang, ``Benchmarking of
  constant power generation strategies for single-phase grid-connected
  photovoltaic systems,'' \emph{IEEE Transactions on Industry Applications},
  vol.~54, no.~1, pp. 447--457, 2017.

\bibitem{tafti2020extended}
H.~D. Tafti, G.~Konstantinou, C.~D. Townsend, G.~G. Farivar, A.~Sangwongwanich,
  Y.~Yang, J.~Pou, and F.~Blaabjerg, ``Extended functionalities of photovoltaic
  systems with flexible power point tracking: recent advances,'' \emph{IEEE
  Transactions on Power Electronics}, vol.~35, no.~9, pp. 9342--9356, 2020.

\bibitem{tafti2018adaptive}
H.~D. Tafti, A.~Sangwongwanich, Y.~Yang, J.~Pou, G.~Konstantinou, and
  F.~Blaabjerg, ``An adaptive control scheme for flexible power point tracking
  in photovoltaic systems,'' \emph{IEEE Transactions on Power Electronics},
  vol.~34, no.~6, pp. 5451--5463, 2018.

\bibitem{paduani2020maximum}
V.~Paduani, L.~Song, B.~Xu, and N.~Lu, ``Maximum power reference tracking
  algorithm for power curtailment of photovoltaic systems,'' in \emph{2021 IEEE
  Power \& Energy Society General Meeting (PESGM)}.\hskip 1em plus 0.5em minus
  0.4em\relax IEEE, 2021, pp. 1--5.

\bibitem{gevorgian2016advanced}
V.~Gevorgian and B.~O'Neill, ``Advanced grid-friendly controls demonstration
  project for utility-scale {PV} power plants,'' National Renewable Energy
  Lab.(NREL), Golden, CO (United States), Tech. Rep., 2016.

\bibitem{sangwongwanich2017delta}
A.~Sangwongwanich, Y.~Yang, F.~Blaabjerg, and D.~Sera, ``Delta power control
  strategy for multistring grid-connected {PV} inverters,'' \emph{IEEE
  Transactions on Industry Applications}, vol.~53, no.~4, pp. 3862--3870, 2017.

\bibitem{gevorgian2019highly}
V.~Gevorgian, ``Highly accurate method for real-time active power reserve
  estimation for utility-scale photovoltaic power plants,'' National Renewable
  Energy Lab.(NREL), Golden, CO (United States), Tech. Rep., 2019.

\bibitem{batzelis2017power}
E.~I. Batzelis, G.~E. Kampitsis, and S.~A. Papathanassiou, ``Power reserves
  control for {PV} systems with real-time mpp estimation via curve fitting,''
  \emph{IEEE Transactions on Sustainable Energy}, vol.~8, no.~3, pp.
  1269--1280, 2017.

\bibitem{batzelis2020mpp}
E.~I. Batzelis, A.~Junyent-Ferre, and B.~C. Pal, ``{MPP} estimation of {PV}
  systems keeping power reserves under fast irradiance changes,'' in \emph{IEEE
  Power \& Energy Society General Meeting {(PESGM)}}.\hskip 1em plus 0.5em
  minus 0.4em\relax IEEE, 2020, pp. 1--5.

\bibitem{zhu2020high}
Y.~Zhu, H.~Wen, G.~Chu, Y.~Hu, X.~Li, and J.~Ma, ``High-performance
  photovoltaic constant power generation control with rapid maximum power point
  estimation,'' \emph{IEEE Transactions on Industry Applications}, vol.~57,
  no.~1, pp. 714--729, 2020.

\bibitem{li2018novel}
X.~Li, H.~Wen, Y.~Zhu, L.~Jiang, Y.~Hu, and W.~Xiao, ``A novel sensorless
  photovoltaic power reserve control with simple real-time mpp estimation,''
  \emph{IEEE Transactions on Power Electronics}, vol.~34, no.~8, pp.
  7521--7531, 2018.

\bibitem{villalva2009comprehensive}
M.~G. Villalva, J.~R. Gazoli, and E.~Ruppert~Filho, ``Comprehensive approach to
  modeling and simulation of photovoltaic arrays,'' \emph{IEEE Transactions on
  power electronics}, vol.~24, no.~5, pp. 1198--1208, 2009.

\bibitem{batzelis2017simple}
E.~I. Batzelis, ``Simple {PV} performance equations theoretically well founded
  on the single-diode model,'' \emph{IEEE Journal of Photovoltaics}, vol.~7,
  no.~5, pp. 1400--1409, 2017.

\bibitem{moler2021lambert}
C.~Mole, ``The {Lambert W} function,''
  \url{https://www.mathworks.com/matlabcentral/fileexchange/43419-the-lambert-w-function},
  {MATLAB} Central File Exchange. Retrieved: December 23, 2020.

\bibitem{gavin2019levenberg}
H.~P. Gavin, ``The {L}evenberg-{M}arquardt algorithm for nonlinear least
  squares curve-fitting problems,'' \emph{Department of Civil and Environmental
  Engineering, Duke University http://people. duke. edu/\~{} hpgavin/ce281/lm.
  pdf}, pp. 1--19, 2019.

\bibitem{marion2014new}
B.~Marion, A.~Anderberg, C.~Deline, J.~del Cueto, M.~Muller, G.~Perrin,
  J.~Rodriguez, S.~Rummel, T.~J. Silverman, F.~Vignola \emph{et~al.}, ``New
  data set for validating pv module performance models,'' in \emph{2014 IEEE
  40th Photovoltaic Specialist Conference (PVSC)}.\hskip 1em plus 0.5em minus
  0.4em\relax IEEE, 2014, pp. 1362--1366.

\bibitem{EPRIdata}
EPRI, ``{Distributed {PV} Monitoring and Feeder Analysis},''
  \url{https://dpv.epri.com/measurement_data.html}, accessed: 2020-08-10.

\bibitem{marcos2011irradiance}
J.~Marcos, L.~Marroyo, E.~Lorenzo, D.~Alvira, and E.~Izco, ``From irradiance to
  output power fluctuations: the pv plant as a low pass filter,''
  \emph{Progress in Photovoltaics: Research and Applications}, vol.~19, no.~5,
  pp. 505--510, 2011.

\bibitem{nielsen1999damping}
H.~B. Nielsen \emph{et~al.}, \emph{Damping parameter in Marquardt's
  method}.\hskip 1em plus 0.5em minus 0.4em\relax IMM, 1999.

\bibitem{long2020diesel}
Q.~Long, H.~Yu, F.~Xie, N.~Xie, and D.~Lubkeman, ``Diesel generator model
  parameterization for microgrid simulation using hybrid box-constrained
  levenberg-marquardt algorithm,'' \emph{IEEE Transactions on Smart Grid},
  vol.~{12}, no.~2, pp. 943--952, 2020.

\bibitem{batzelis2015method}
E.~I. Batzelis and S.~A. Papathanassiou, ``A method for the analytical
  extraction of the single-diode {PV} model parameters,'' \emph{IEEE
  Transactions on Sustainable Energy}, vol.~7, no.~2, pp. 504--512, 2015.

\bibitem{saloux2011explicit}
E.~Saloux, A.~Teyssedou, and M.~Sorin, ``Explicit model of photovoltaic panels
  to determine voltages and currents at the maximum power point,'' \emph{Solar
  energy}, vol.~85, no.~5, pp. 713--722, 2011.

\bibitem{TIsensor}
\relax Mouser~Electronics, ``Texas {I}nstruments {AMC1303x} high-precision
  delta-sigma modulators",''
  \url{https://www.mouser.it/new/TexasInstruments/ti-amc1303-modulators/},
  retrieved: March 31, 2021.

\bibitem{edstrom2012modeling}
F.~Edstrom, J.~Rosenlind, P.~Hilber, and L.~Soder, ``Modeling impact of cold
  load pickup on transformer aging using ornstein-uhlenbeck process,''
  \emph{IEEE Transactions on power delivery}, vol.~27, no.~2, pp. 590--595,
  2012.

\bibitem{sangwongwanich2018analysis}
A.~Sangwongwanich, Y.~Yang, D.~Sera, H.~Soltani, and F.~Blaabjerg, ``Analysis
  and modeling of interharmonics from grid-connected photovoltaic systems,''
  \emph{IEEE Transactions on power electronics}, vol.~33, no.~10, pp.
  8353--8364, 2018.

\bibitem{pattabiraman2018impact}
D.~Pattabiraman, J.~Tan, V.~Gevorgian, A.~Hoke, C.~Antonio, and D.~Arakawa,
  ``Impact of frequency-watt control on the dynamics of a high der penetration
  power system,'' in \emph{2018 IEEE Power \& Energy Society General Meeting
  (PESGM)}.\hskip 1em plus 0.5em minus 0.4em\relax IEEE, 2018, pp. 1--5.

\bibitem{paduani2019small}
V.~Paduani, M.~Kabalan, and P.~Singh, ``Small-signal stability of
  islanded-microgrids with dc side dynamics of inverters and saturation of
  current controllers,'' in \emph{2019 IEEE Power \& Energy Society General
  Meeting (PESGM)}.\hskip 1em plus 0.5em minus 0.4em\relax IEEE, 2019, pp.
  1--5.

\bibitem{instruments2008tms320c28x}
\relax Texas~Instruments, ``{TMS320C28x} floating point unit and instruction
  set reference guide,'' 2008.

\end{thebibliography}

\vspace*{10cm}

\begin{IEEEbiography}[{\includegraphics[width=1in,height=1.25in,clip,keepaspectratio]{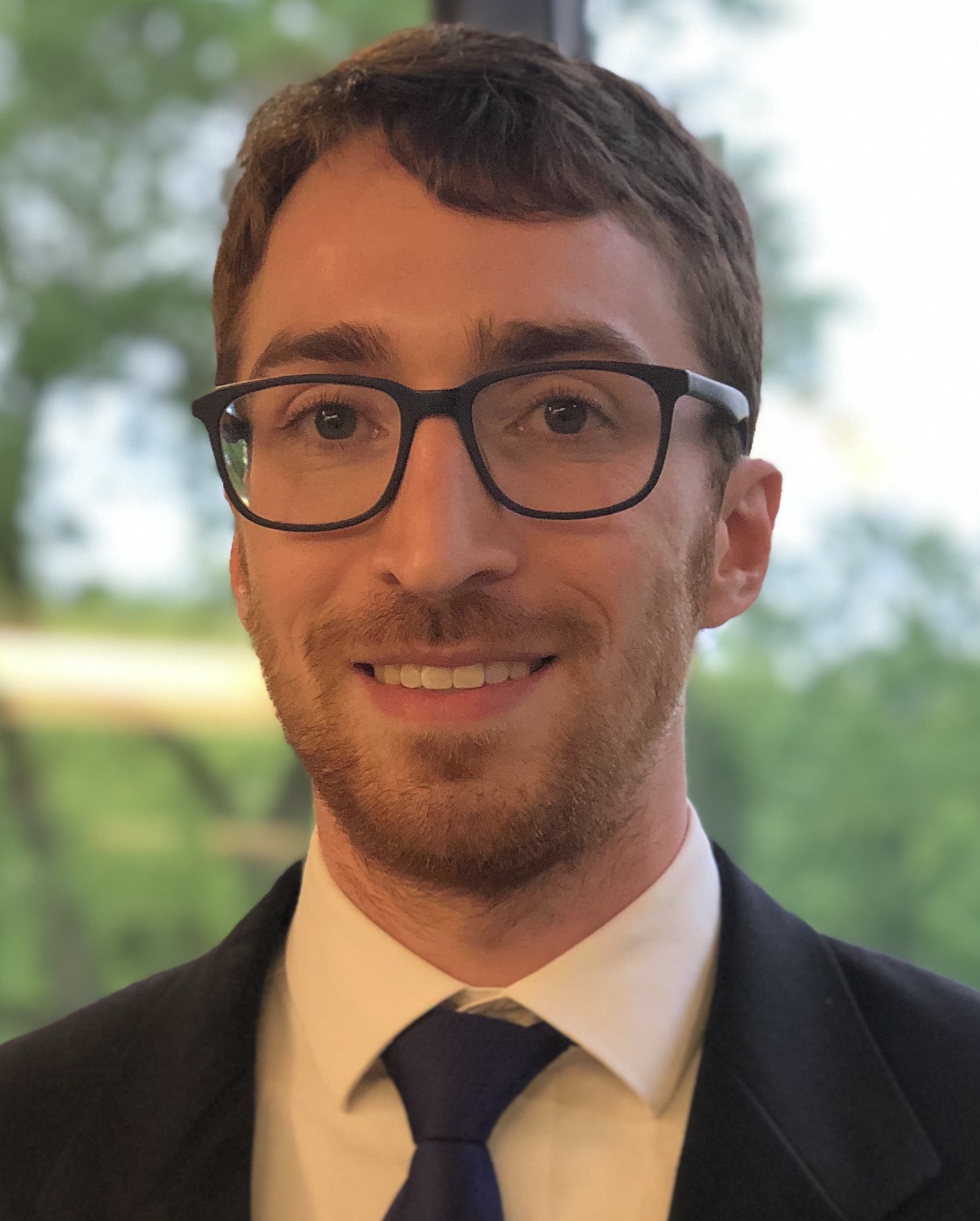}}]{Victor Paduani}
(Student Member, IEEE) received the B.S. degree in electrical engineering from Federal University of Santa Catarina, Florianopolis, Brazil, in 2017, and the M.S. degree in electrical engineering from Villanova University, Villanova, USA, in 2019. He is currently pursuing the Ph.D. degree in electrical engineering with the Future Renewable Electric Energy Delivery and Management (FREEDM) Systems Center, North Carolina State University, Raleigh, USA. His research interests include modeling distributed energy resources, microgrid controller design, and developing real-time, hardware-in-the-loop test systems that co-simulate power electronic systems, power system networks, and energy management systems.
\end{IEEEbiography}
\vskip 0pt plus -1fil
\begin{IEEEbiography}[{\includegraphics[width=1in,height=1.25in,clip,keepaspectratio]{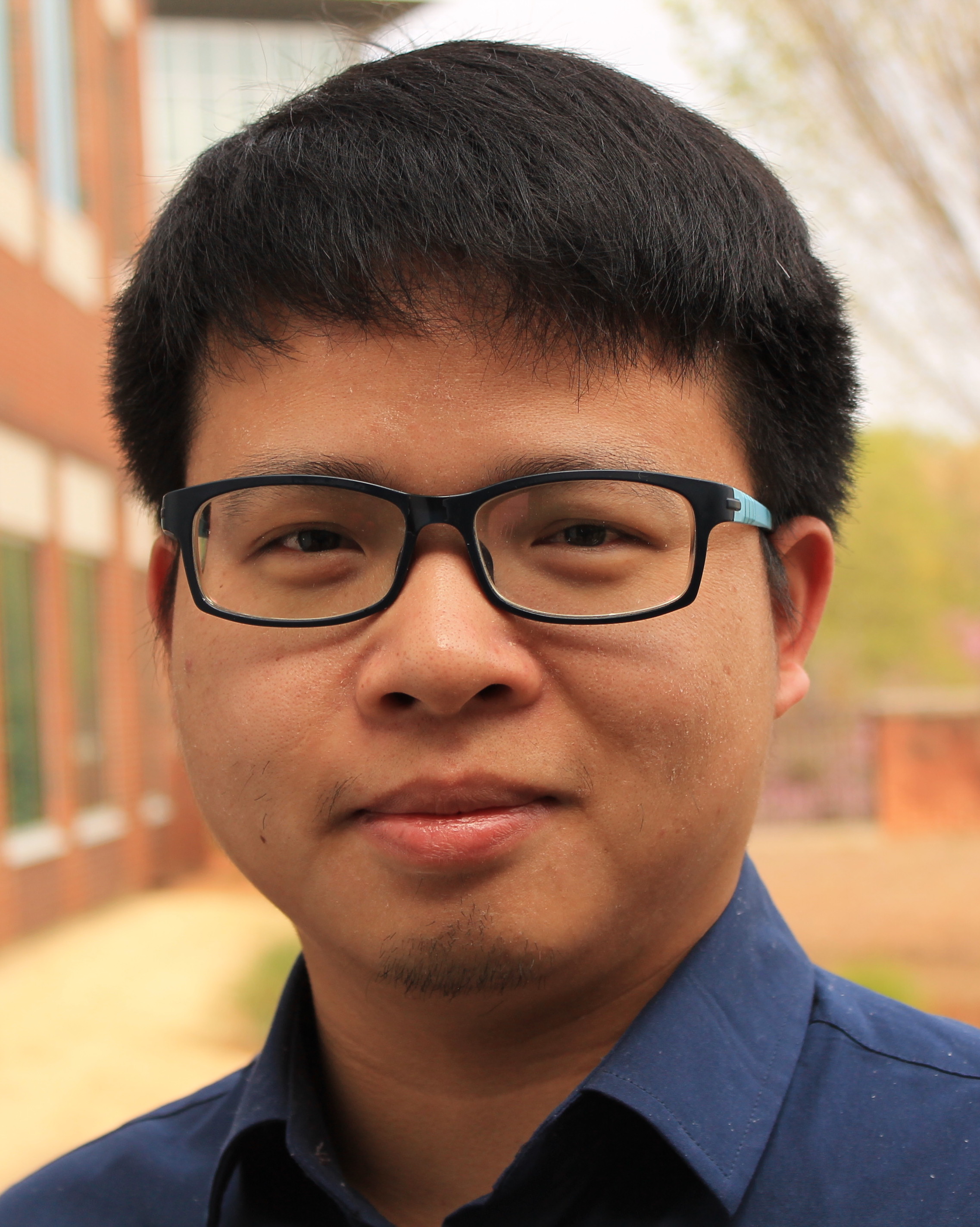}}]{Hui Yu}
(Member, IEEE) completed the bachelor's and master's degrees from Huazhong University of Science and Technology, Wuhan, China, in 2013 and 2016, respectively, and the Ph.D. degree from North Carolina State University, Raleigh, NC, USA, in 2020, all in electrical engineering. He is currently a Postdoctral Research Scholar  with the Future Renewable Electric Energy Delivery and Management Systems Center, North Carolina State University. His research interests include power electronics converter modeling, design, and control in microgrids and energy storage systems.
\end{IEEEbiography}

\vskip 0pt plus -1fil
\begin{IEEEbiography}
[{\includegraphics[width=1in,height=1.25in,clip,keepaspectratio]{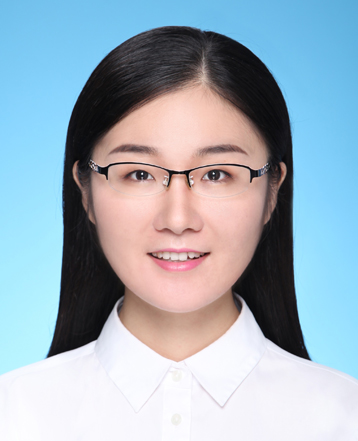}}]{Bei Xu}
(Student Member, IEEE) received the bachelor’s and master’s degrees in electrical engineering from Beijing Jiaotong University, Beijing, China, in 2016 and 2019, respectively. She is currently a Ph.D. candidate in electrical engineering with the Future Renewable Electric Energy Delivery and Management (FREEDM) Systems Center, North Carolina State University, Raleigh, NC, USA. Her research interests include developing grid-following and grid-forming functions for battery energy storage systems. 
\end{IEEEbiography}
\vskip 0pt plus -1fil
\begin{IEEEbiography}[{\includegraphics[width=1in,height=1.25in]{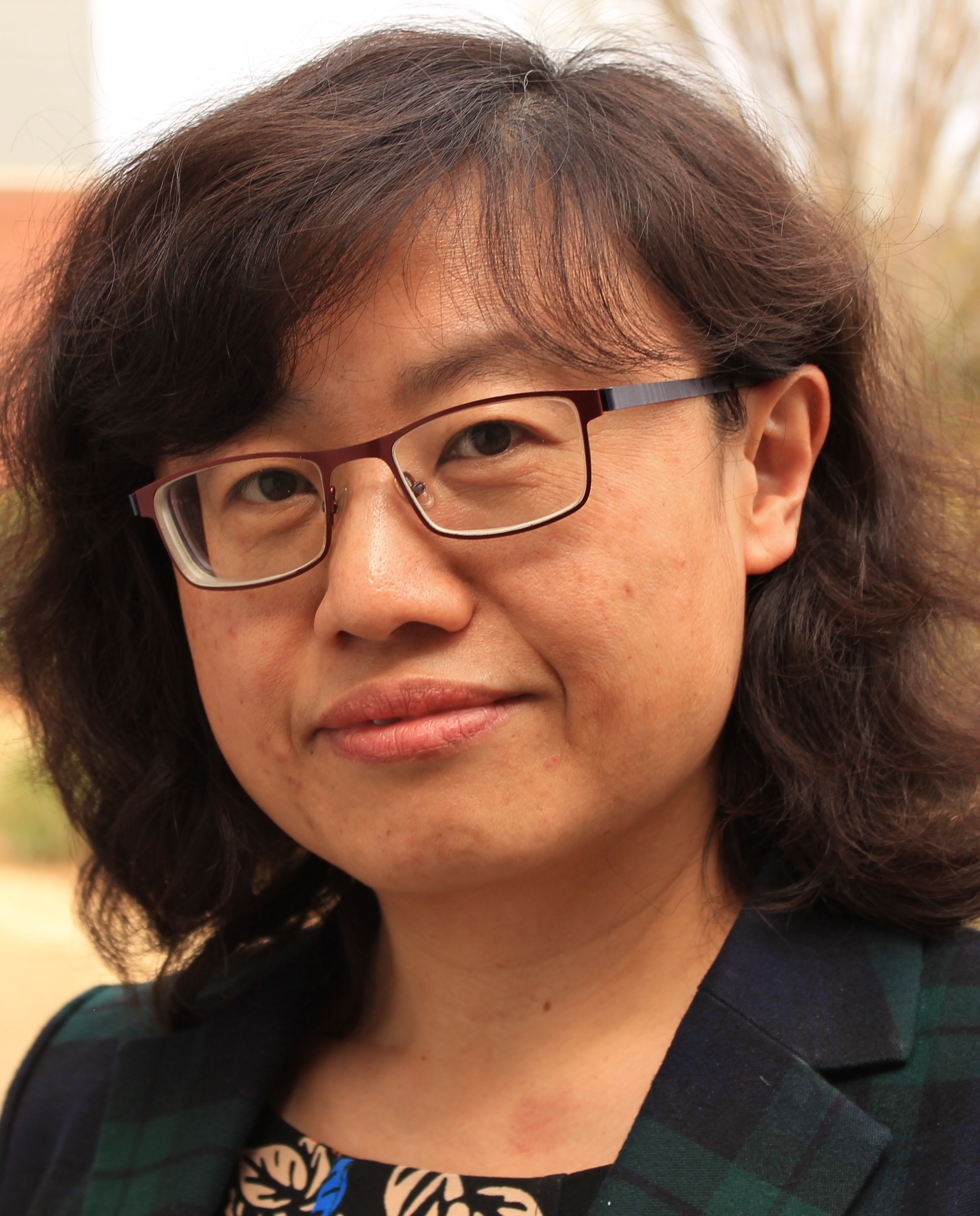}}]{Ning Lu}
(Fellow, IEEE) received the B.S. degree in electrical engineering from the Harbin Institute of Technology, Harbin, China, in 1993, and the M.S. and Ph.D. degrees in electric power engineering from Rensselaer Polytechnic Institute, Troy, NY, USA, in 1999 and 2002, respectively. She is a professor with the Electrical and Computer Engineering Department, North Carolina State University. Her research interests include modeling and control of distributed energy resources, microgrid energy management, and applying machine learning methods in energy forecasting, modeling, and control.
\end{IEEEbiography}

\end{document}